\documentclass[aps,prl,showpacs,twocolumn,superscriptaddress]{revtex4-1}
\usepackage{amsmath}
\usepackage{amsfonts}
\usepackage{amssymb}
\usepackage{mathtools}
\usepackage{epsfig}
\usepackage[autostyle]{csquotes}
\usepackage{braket}
\usepackage{graphicx}
\usepackage{bbold}

\usepackage{hyphenat}
\usepackage{hyperref}
\usepackage[all]{hypcap}
\usepackage{placeins}	
\usepackage{color}
\usepackage{soul}
\usepackage{tikz}
\usetikzlibrary{
  arrows.meta,
  positioning,
  calc,
  decorations.pathreplacing,
  shapes.geometric
}

\newcommand{\Z}{\mathbb{Z}}
\newcommand{\ZZ}{\mathbb{Z}_2}
\newcommand{\ZZZ}{\mathbb{Z}_2\times\mathbb{Z}_2\times\mathbb{Z}_2}
\newcommand{\R}{\mathbb{R}}

\newcommand{\up}{\uparrow}
\newcommand{\down}{\downarrow}

\newcommand{\blue}[1]{\textcolor{dunkelblau}{#1}}
\newcommand{\green}[1]{\textcolor{FAUdunkelgruen}{#1}}
\newcommand{\red}[1]{\textcolor{dunkelrot}{#1}}
\definecolor{rosa}{HTML}{FF88D5}
\definecolor{lila}{HTML}{82218B}
\definecolor{dunkelblau}{HTML}{03468F}
\definecolor{mittelblau}{HTML}{5AA0E5}
\definecolor{FAUgruen}{RGB}{123,183,37}
\definecolor{FAUdunkelgruen}{RGB}{38,97,65}
\definecolor{dunkelrot}{HTML}{92002f}
\definecolor{hellgrau}{HTML}{878787}

\DeclareRobustCommand{\trirobust}{
  \mathord{\tikz[baseline=-0.2ex]
      \node[
        regular polygon,
        regular polygon sides=3,
        draw,
        minimum size=1ex,
        inner sep=0pt
      ] {}; }}

\newcommand{\tri}{
  \mathord{\tikz[baseline=-0.2ex]
      \node[
        regular polygon,
        regular polygon sides=3,
        draw,
        minimum size=1ex,
        inner sep=0pt
      ] {}; }}

\newcommand{\intri}{
  \mathord{\tikz[baseline=-0.2ex]
      \node[
        regular polygon,
        regular polygon sides=3,
        draw,
        minimum size=1.3ex,
        inner sep=0pt
      ] {}; }}
      
\DeclareRobustCommand{\hexrobust}{
  \mathord{\tikz[baseline=-0.2ex]
      \node[
        regular polygon,
        regular polygon sides=6,
        rotate=30,
        draw,
        minimum size=1ex,
        inner sep=0pt
      ] {}; }}

\newcommand{\hex}{
  \mathord{\tikz[baseline=-0.2ex]
      \node[
        regular polygon,
        regular polygon sides=6,
        rotate=30,
        draw,
        minimum size=1ex,
        inner sep=0pt
      ] {}; }}
      
 \newcommand{\hexin}{
  \mathord{\tikz[baseline=-0.4ex]
      \node[
        regular polygon,
        regular polygon sides=6,
        rotate=30,
        draw,
        minimum size=1ex,
        inner sep=0pt
      ] {}; }}  
      
\newcommand{\inhex}{
  \mathord{\tikz[baseline=-0.25ex]
      \node[
        regular polygon,
        regular polygon sides=6,
        rotate=30,
        draw,
        minimum size=1.3ex,
        inner sep=0pt
      ] {}; }}

\newcommand{\ininhex}{
  \mathord{\tikz[baseline=-0.55ex]
      \node[
        regular polygon,
        regular polygon sides=6,
        rotate=30,
        draw,
        minimum size=1.9ex,
        inner sep=0pt
      ] {}; }}
      
\newcommand{\dodec}{
  \mathord{\tikz[baseline=-0.4ex]
      \node[
        regular polygon,
        regular polygon sides=12,
        draw,
        minimum size=1.3ex,
        inner sep=0pt
      ] {}; }}
      
\newcommand{\centertriangle}[1][0.18]{
  \mathbin{
    \begin{tikzpicture}[baseline=-0.4ex, scale=0.15, line cap=round]
      \coordinate (A) at (90:1);
      \coordinate (B) at (210:1);
      \coordinate (C) at (330:1);
      \fill (A) circle[radius=0.15];
      \fill (B) circle[radius=0.15];
      \fill (C) circle[radius=0.15];
      \fill (0,0) circle[radius=0.15];
      \draw (A) -- (B) -- (C) -- cycle;
      \draw[thin] (A) -- (0,0) (B) -- (0,0) (C) -- (0,0);
    \end{tikzpicture}
  }}

\newcommand{\trianglewithoutcenter}[1][0.18]{
  \mathbin{
    \begin{tikzpicture}[baseline=-0.4ex, scale=0.15, line cap=round]
      \coordinate (A) at (90:1);
      \coordinate (B) at (210:1);
      \coordinate (C) at (330:1);
      \fill (A) circle[radius=0.15];
      \fill (B) circle[radius=0.15];
      \fill (C) circle[radius=0.15];
      \draw (A) -- (B) -- (C) -- cycle;
    \end{tikzpicture}
  }}
\newcommand{\centerhexagon}[1][0.18]{
  \mathbin{
    \begin{tikzpicture}[baseline=-0.4ex, scale=0.15, line cap=round]
      \foreach \i in {0,...,5} {
        \coordinate (H\i) at ({90 - 60*\i}:1);
      }
      \draw[thin]
        (H0) -- (H1) -- (H2) -- (H3) -- (H4) -- (H5) -- cycle;
      \foreach \i in {0,...,5} {
        \fill (H\i) circle[radius=0.15];
      }
    \end{tikzpicture}}}
\newcommand{\hexstatePlatzhalter}[1]{\mathbin{
    \tikz[baseline=-0.5ex, scale=0.15, line cap=round]{\foreach \i in {0,...,5} {\coordinate (H\i) at ({90 - 60*\i}:1);}
        \draw[thick, hellgrau] (H0) -- (H1) -- (H2) -- (H3) -- (H4) -- (H5) -- cycle;
        \foreach \i in {0,...,5} {\fill[hellgrau] (H\i) circle[radius=0.25];}}}}
\newcommand{\hexstateeven}{\mathbin{
    \tikz[baseline=-0.5ex, scale=0.15, line cap=round]{\foreach \i in {0,...,5} {\coordinate (H\i) at ({90 - 60*\i}:1);}
        \draw[thick,hellgrau] (H0) -- (H1) -- (H2) -- (H3) -- (H4) -- (H5) -- cycle;
        \foreach \i in {0,...,5} {\fill[hellgrau] (H\i) circle[radius=0.25];}}_{\text{even}}}}
\newcommand{\hexstateuneven}{\mathbin{
    \tikz[baseline=-0.5ex, scale=0.15, line cap=round]{\foreach \i in {0,...,5} {\coordinate (H\i) at ({90 - 60*\i}:1);}
        \draw[thick,hellgrau] (H0) -- (H1) -- (H2) -- (H3) -- (H4) -- (H5) -- cycle;
        \foreach \i in {0,...,5} { \fill[hellgrau] (H\i) circle[radius=0.25];}}_{\text{uneven}}}}
\newcommand{\sixtriangles}[1]{
  \mathbin{
    \tikz[baseline=-0.4ex, scale=0.1, line cap=round]{

    \def\a{1}   % Gitterkonstante
    \def\Nx{2}  % Anzahl Zellen in x-Richtung
    \def\Ny{1}  % Anzahl Zellen in y-Richtung
    \def\h{sqrt(3)/2*\a}
    \def\r{sqrt(3)/6*\a}
    \def\R{sqrt(3)/3*\a}
    
    \coordinate (a0) at (0,0);
    \coordinate (a1) at (0,{-2*\h});
    \coordinate (a2) at ({3/2*\a}, {-\h});
    \coordinate (a3) at ({3/2*\a}, {\h});
    \coordinate (a4) at (0,{2*\h});
    \coordinate (a5) at ({-3/2*\a}, {\h});
    \coordinate (a6) at ({-3/2*\a}, {-\h});
    
    \coordinate (b1) at (0,{-\h});
    \coordinate (b2) at ({3/4*\a},{-1/2*\h});
    \coordinate (b3) at ({3/4*\a}, {1/2*\h});
    \coordinate (b4) at (0, {\h});
    \coordinate (b5) at ({-3/4*\a}, {1/2*\h});
    \coordinate (b6) at ({-3/4*\a},{-1/2*\h});
    \coordinate (b7) at ({\a/4},{-3/2*\h});
    \coordinate (b8) at ({\a},{-\h});
    \coordinate (b9) at ({\a+\a/4}, {-1/2*\h});
    \coordinate (b10) at ({\a+\a/4}, {1/2*\h});
    \coordinate (b11) at ({\a},{\h});
    \coordinate (b12) at ({\a/4},{3/2*\h});
    \coordinate (b13) at ({-\a/4},{3/2*\h});
    \coordinate (b14) at ({-\a},{\h});
    \coordinate (b15) at ({-\a-\a/4}, {1/2*\h});
    \coordinate (b16) at ({-\a-\a/4}, {-1/2*\h});
    \coordinate (b17) at ({-\a},{-\h});
    \coordinate (b18) at ({-\a/4},{-3/2*\h});
    
        \draw[hellgrau] (a2) -- (a4) -- (a6) -- (a2);
        \draw[hellgrau] (a1) -- (a3) -- (a5) -- (a1);
    
    \begin{pgfonlayer}{main}
        \foreach \c in {b1,b2,b3,b4,b5,b6,b7,b8,b9,b10,b11,b12,b13,b14,b15,b16,b17,b18}{
        \fill[hellgrau] ($(\c)$) circle;}
        
    \end{pgfonlayer}
        \draw[lila] (b1) -- (b7) -- (b18) -- (b1);
        \draw[lila] (b2) -- (b8) -- (b9) -- (b2);
        \draw[lila] (b3) -- (b10) -- (b11) -- (b3);
        \draw[lila] (b4) -- (b12) -- (b13) -- (b4);
        \draw[lila] (b5) -- (b14) -- (b15) -- (b5);
        \draw[lila] (b6) -- (b16) -- (b17) -- (b6);
        }
    }
}

\begin{document}

\title{Unconventional $\ZZZ$ topological order in the kagome XY toric code}

\author{Constanze K\"olbl}
\email{constanze.koelbl@fau.de}
\affiliation{Department Physik, Staudtstra{\ss}e 7, Universit\"at Erlangen-N\"urnberg, D-91058 Erlangen, Germany}
\author{Maximilian Vieweg}
\email{max.vieweg@fau.de}
\affiliation{Department Physik, Staudtstra{\ss}e 7, Universit\"at Erlangen-N\"urnberg, D-91058 Erlangen, Germany}

\author{Kai Phillip Schmidt}
\email{kai.phillip.schmidt@fau.de}
\affiliation{Department Physik, Staudtstra{\ss}e 7, Universit\"at Erlangen-N\"urnberg, D-91058 Erlangen, Germany}

\begin{abstract}
We investigate the quantum phase diagram of the XY toric code (XYTC) on the kagome lattice consisting of $XY$ hexagonal and triangular plaquette operators and conserved star operators. 
We demonstrate analytically by exploiting an exact local $\ZZ$-symmetry on dodecagons that the kagome XYTC realizes a $\ZZZ$ topologically ordered phase in the limit of large hexagonal \mbox{plaquette} operators. 
This unconventional topological phase involves 64 quasi-particles - Abelian anyons - with restricted mobility on three colored sublattices
establishing a quantum dimension of eight. 
The large number of topological superselection sectors originates from six independent Wilson loop operators, acting as emergent one-form symmetries. 
The resulting anyon structure is richer than that of established topological codes like the toric and the color code. A corresponding CSS topological stabilizer code is formulated, opening novel possibilities for the encoding and manipulation of topological quantum information.
\end{abstract}

\maketitle
%Introduction
%%%%%%%%%%%%%%%%%%%%%%%%%%%%%%%%%%%%%%%%%%%%%%%%%%%%%%%%%%%%%%%%%%%%%%%%%%%%%%%%%%%%%%%%%%%%
Highly entangled quantum many-body states constitute a key resource for quantum information processing. In this context, topologically ordered phases  \cite{Wen_1989,Wen_1990,Wen_2004} represent a particularly remarkable form of quantum matter, characterized by long-range entanglement and fractionalized anyonic excitations \cite{Leinaas_1977,Wilczek_1982}. Their robustness against local perturbations enables the nonlocal storage and manipulation of quantum information, making them a leading candidate for fault-tolerant quantum memories and topological quantum computation \cite{Kitaev_2003,Nayak_2008}. Understanding, preparing, and characterizing such states has therefore become a major objective in condensed-matter physics and quantum information science. Experimental investigations span a wide range of platforms, from fractional quantum Hall systems \cite{Laughlin_1983,Tsui_1982} and frustrated magnets \cite{Balents_2010,Jackeli_2010,Singh_2012,Plumb_2014,Banerjee_2017,Savary_2017,Banerjee_2018} to quantum simulators based on trapped ions, photons, and nuclear magnetic resonance techniques \cite{Micheli_2006,Du_2007,Paredes_2008,Feng_2012,Peng_2014,Sameti_2017}.

A paradigmatic realization of topological order is Kitaev’s exactly solvable 2D $\mathbb{Z}_2$ toric code \cite{Kitaev_2003}. Beyond providing a minimal model of long-range entanglement and anyonic quasiparticles, the toric code established a conceptual foundation for topological quantum memories and modern quantum error-correcting codes. Its ground state exhibits $\mathbb{Z}_2$ topological order, while its elementary excitations are mutually Abelian anyons. Owing to its analytical tractability and relevance for fault-tolerant quantum information processing, the model has served as a benchmark for extensive studies of the stability of topological order against quantum and thermal perturbations \cite{Trebst_2007,Hamma_2008_b,Yu_2008,Vidal_2009,Vidal_2011,Dusuel_2009,Tupitsyn_2010,Wu_2012,Dusuel_2011,Schmidt_2013,Morampudi_2014,Zhang_2017,Vanderstraeten_2017,Alicki_2009,Castelnovo_2007,Nussinov_2009_b,Kott2024}. Motivated by the search for more versatile topological codes and quantum-information architectures, numerous generalizations have been proposed, ranging from models based on alternative gauge groups \cite{Kitaev_2003,Bais_2009,WOOTTON20112307,Bullock2007,Schulz_2012} to constructions exploiting combinatorial gauge symmetries \cite{Chamon_2020}, which have been shown to be compatible with superconducting-circuit implementations \cite{Chamon_2021}.
Recently, the symmetry-deformed $U(1)$ toric code (U1TC) \cite{Wu_2023,Qiao_2025,Vieweg_2026} and the XY toric code (XYTC) \cite{Vieweg_2025,Qiao_2025} on the square lattice have uncovered unexpected links between topological order and fracton phenomena. While the XYTC supports fractonic excitations with restricted mobility protected by subsystem symmetries, the U1TC has a non-topological ground state with confined fractons \cite{Vieweg_2026,Qiao_2025}. 

%Figure 1 - Kagome XYTC
%%%%%%%%%%%%%%%%%%%%%%%%%%%%%%%%%%%%%%%%%%%%%%%%%%%%%%%%%%%%%%%%%%%%%%%%%%%%%%%%%%%%%%%%%%%%
\begin{figure}[t]
       \centering
        \includegraphics[width=0.9\columnwidth]{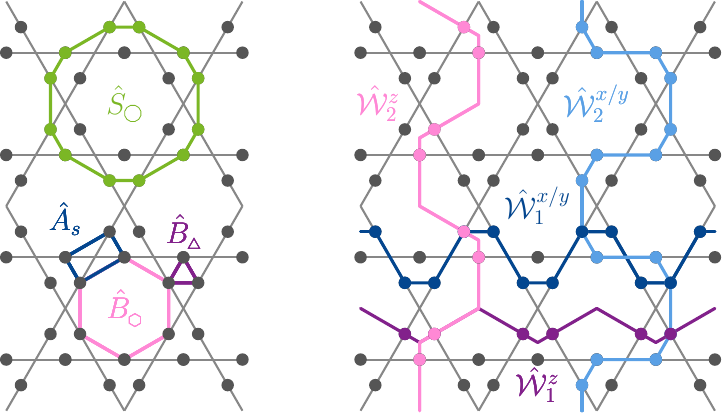}
        \caption[\textit{Left}: XYTC on the kagome lattice. Star operators \(\hat{A}_s\) are depicted in dark blue, and two different kinds of xy-plaquette operators 
        \(\hat{B}_{tri}\) (purple) and \(\hat{B}_{hex}\) (pink) 
        are distinguished. Local symmetries are individual star operators and 12-spin operators \(\hat{S}\) along a dodecagon (green).
        \textit{Right}: Wilson loop operators are one-form symmetries of the system with implicit periodic boundary conditions. 
        ]{\textit{Left}: XYTC on the kagome lattice. Star operators \(\hat{A}_s\) are depicted in dark blue, and two different kinds of xy-plaquette operators 
        \(\hat{B}_{\tri}\) (purple) and \(\hat{B}_{\hex}\) (pink) 
        are distinguished. Local symmetries are individual star operators and 12-spin operators \(\hat{S}_{\dodec}\) along dodecagons (green).
        \textit{Right}: Wilson loop operators \(\hat{\mathcal{W}}^{\alpha}_{n}\) are one-form symmetries of the system with implicit periodic boundary conditions. 
        }
       \label{Fig:XYTC}
\end{figure}
%%%%%%%%%%%%%%%%%%%%%%%%%%%%%%%%%%%%%%%%%%%%%%%%%%%%%%%%%%%%%%%%%%%%%%%%%%%%%%%%%%%%%%%%%%%%

In this Letter, we establish that the symmetry-deformed XYTC on the kagome lattice hosts an unconventional $\mathbb{Z}_2 \times \mathbb{Z}_2 \times \mathbb{Z}_2$ topological order characterized by a total quantum dimension of eight and 64 Abelian anyon types.
The resulting anyon structure is considerably richer than that of established topological codes, opening novel possibilities for the encoding and manipulation of topological quantum information.

%Model
%%%%%%%%%%%%%%%%%%%%%%%%%%%%%%%%%%%%%%%%%%%%%%%%%%%%%%%%%%%%%%%%%%%%%%%%%%%%%%%%%%%%%%%%%%%%
{\it{Kagome XYTC:}}
The kagome XYTC is illustrated in Fig.~\ref{Fig:XYTC}. 
Its Hamiltonian is given by 
\begin{align} \label{eq:Hamiltonian_XYTC_kagome}
    &\hat{\mathcal{H}} = -J_{\tri} \sum_{\tri} \hat{B}_{\tri}(\varphi) - J_{\hex}\sum_{\hex} \hat{B}_{\hex}(\varphi) - J_s \sum_s \hat{A}_s
\end{align}
with $J_{i}\geq 0$ for \(i\in \{\intri,\inhex,s\}\), acting on spin \(1/2\) degrees of freedom. Star operators encompass four spins with Pauli \(z\)-interaction $\hat{A}_s = \prod_{i\in s} \sigma_i^z$.
Due to the geometry of the kagome lattice, two different types of plaquette operators \(\hat{B}_{\tri}\) and \(\hat{B}_{\hex}\) are considered, including both Pauli \(x\)- and \(y\)- contributions.
Defining $\hat{B}_{\tri/ \hex}^{(\alpha)}\equiv \prod_{i\in\intri/ \inhex} \sigma_i^\alpha$ with $\alpha\in\{x,y,z\}$ and eigenvalues $b_{\tri/ \hex}^{(\alpha)}=\pm 1$, one has the general form 
\begin{align}
    \hat{B}_{\tri/ \hex}(\varphi) =  \cos{\varphi} \hat{B}_{\tri/ \hex}^{(x)} + \sin{\varphi}\hat{B}_{\tri/ \hex}^{(y)}
\end{align}
for the plaquette operators.
\(\hat{B}_{\hex}\) operators encompass six spins on hexagons, whereas \(\hat{B}_{\tri}\) operators include three-spin interactions on triangles. 
In the following, we set \(J_{\hex}=1-J_{\tri}\,\in[0,1]\) and consider \(\varphi \in \left[0,\frac{\pi}{2}\right]\). 

%Conserved quantities
%%%%%%%%%%%%%%%%%%%%%%%%%%%%%%%%%%%%%%%%%%%%%%%%%%%%%%%%%%%%%%%%%%%%%%%%%%%%%%%%%%%%%%%%%%%%
{\it{Conserved quantities:}} 
The individual star operators $\hat{A}_{s}$ commute with $\hat{\mathcal{H}}$ so that the $a_{s}=\pm 1$ eigenvalues are conserved quantities. In the ground-state manifold, \(a_s = +1\,\forall s\).
This is generally not the case for \(\hat{B}_{\tri}\) and \(\hat{B}_{\hex}\), as they do not commute with $\hat{\mathcal{H}}$ owing to \(\left[\hat{B}_{\tri}, \hat{B}_{\hex} \right] \neq 0\).
Instead, a local symmetry is constituted by 
    \begin{align}
        \hat{S}_{\dodec} = \alpha \prod_{i\in \dodec }\sigma_i^x + \beta \prod_{i\in \dodec }\sigma_i^y \equiv \alpha \hat{S}_{\dodec}^{(x)} + \beta \hat{S}_{\dodec}^{(y)}
    \end{align}
with \(\alpha,\beta \in \R\), involving 12 spins along a dodecagon (see Fig.~\ref{Fig:XYTC}). 
In the symmetry sector \(a_s = +1\,\forall s\), \(\hat{S}_{\dodec}^{(x)}\) and \(\hat{S}_{\dodec}^{(y)}\) are identical, yielding only one conserved quantity per dodecagon.

When imposing periodic boundary conditions, besides local symmetries, independent one-form symmetries are present in the form of Wilson loop operators 
%- closed string operators winding around the torus along non-contractible paths. They can be formulated either in x/y- or z-basis 
    \begin{align}
    \hat{\mathcal{W}}^z_{n} = \prod_{i\in\mathcal{C}_z} \sigma^z_i \, ,\quad \hat{\mathcal{W}}^{x/y}_{n}= \prod_{i\in\mathcal{C}_{x/y}} \sigma^{x/y}_i,
    \end{align}
acting along the non-contractible paths \(\mathcal{C}_z\) or \(\mathcal{C}_{x/y}\) depicted in Fig.~\ref{Fig:XYTC}. Per homology class 
\(n \in \{1,2 \} \) of the torus, there exists one independent Wilson loop, yielding a total of two independent Wilson loop operators with the corresponding conserved eigenvalues \(\pm 1\).

%Figure 2 - Emergent symmetries
%%%%%%%%%%%%%%%%%%%%%%%%%%%%%%%%%%%%%%%%%%%%%%%%%%%%%%%%%%%%%%%%%%%%%%%%%%%%%%%%%%%%%%%%%%%%
\begin{figure}[t]
        \centering
    \includegraphics[width=0.9\columnwidth]{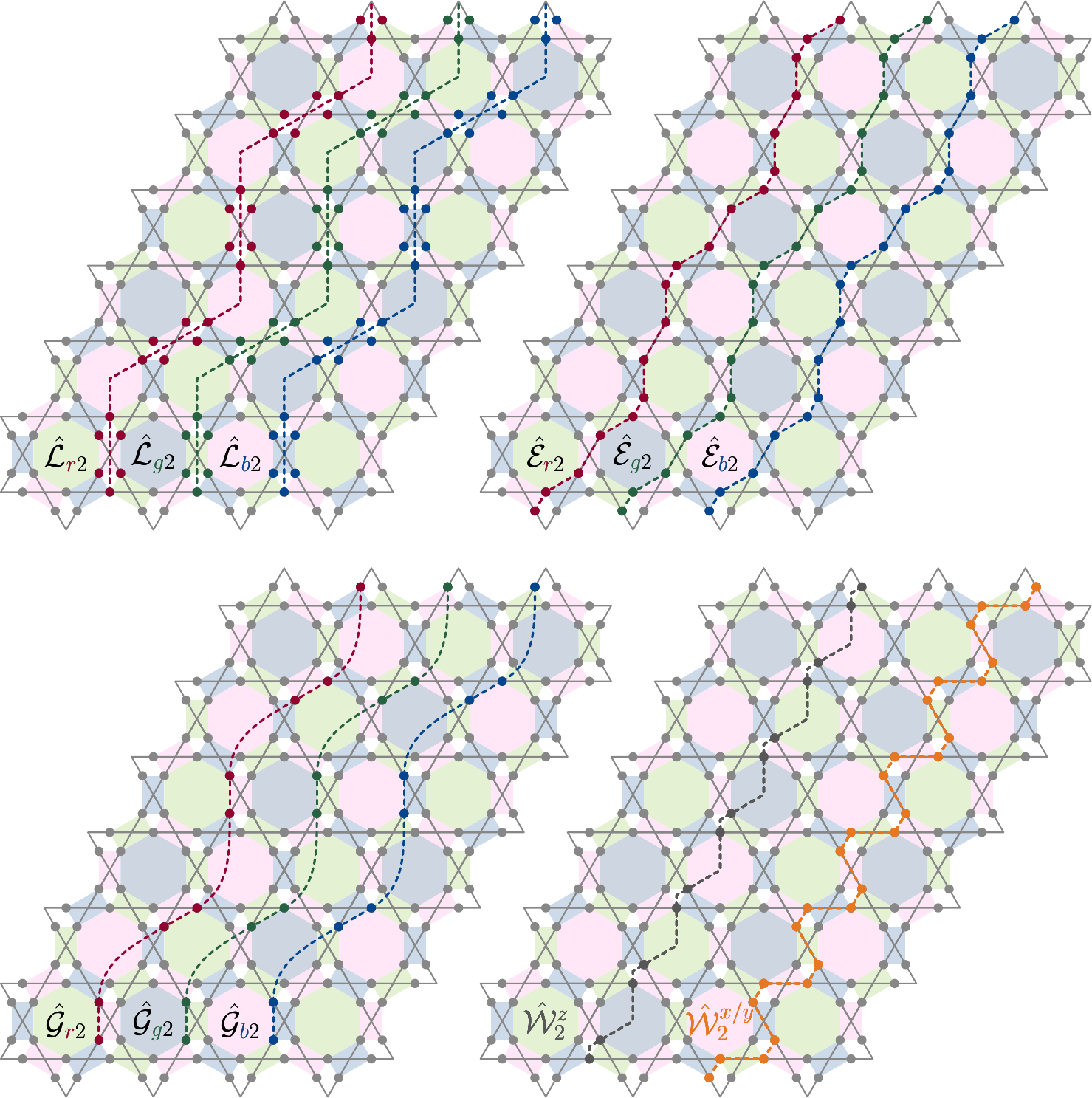}
        \caption{Three different types of emergent symmetries: \(\hat{\mathcal{L}}\) (upper left), \(\hat{\mathcal{E}}\) (upper right) and \(\hat{\mathcal{G}}\) (lower left). Within each category, three colors (red, green, and blue) are distinguished owing to the restricted shifting behavior through the action of local symmetries. 
        \textit{Lower right}: Exact Wilson loop operators \(\hat{\mathcal{W}}^z_{2}\) (black) and \(\hat{\mathcal{W}}^{x/y}_{2}\) (orange). Implicit periodic boundary conditions, connecting top and bottom rows, are assumed.}
        \label{Fig::Emergent_symmetries}
\end{figure}
%%%%%%%%%%%%%%%%%%%%%%%%%%%%%%%%%%%%%%%%%%%%%%%%%%%%%%%%%%%%%%%%%%%%%%%%%%%%%%%%%%%%%%%%%%%%
%Z2 x Z2 x Z2 topo
%%%%%%%%%%%%%%%%%%%%%%%%%%%%%%%%%%%%%%%%%%%%%%%%%%%%%%%%%%%%%%%%%%%%%%%%%%%%%%%%%%%%%%%%%%%%
{\it $\ZZZ$ topological order}: 
We demonstrate analytically that the kagome XYTC exhibits $\ZZZ$ topological order in the limit of small \(J_{\tri}\). To this end, we first consider the exact limit  \mbox{\(J_{\tri} = 0\)} of isolated hexagonal plaquettes and analyze its highly degenerate ground-state manifold. Then we derive an effective low-energy model by degenerate perturbation theory.

Per hexagon, there exist 16 local ground states for \(J_{\tri} = 0\), corresponding to the eigenvalue \(\cos{\varphi}+\sin{\varphi}\) with $b_{\hex}^{(x)}=b_{\hex}^{(y)}=+1$. Enforcing \(a_s=1\,\forall s\) constraints in the low-energy regime reduces the number of global ground-state configurations. Solving these constraints in the case of periodic boundary conditions leads to an extensively degenerate ground-state manifold (\(\#\mathrm{GS} = 2^{N_{\hex}+3}\) with $N_{\hex}=N/6$ the total number of hexagons and \(N\) the total number of spins).

This ground-state degeneracy can be explained through additional dependencies of \(\hat{A}_s\) operators. We partition the lattice into three sublattices \(\Lambda_{\green{g}}\), \(\Lambda_{\blue{b}}\), and \(\Lambda_{\red{r}}\) with colors $c\in\{\green{g},\blue{b},\red{r}\}$ (see Fig.~\ref{Fig::Emergent_symmetries}). 
We have the following exact relation
\begin{align}
 \prod_{s\in \Lambda_{\green{g}}} \hat{A}_s \prod_{\hexin \in \Lambda_{\blue{b}}} \hat{B}_{\hex}^{(z)} \prod_{\hexin \in \Lambda_{\red{r}}} \hat{B}_{\hex}^{(z)} = \mathbb{1} 
\end{align}
and the same for any permutation of the three colors.
For the specific case $J_{\tri}=0$ and in the ground-state manifold, one has $b_{\hex}^{(z)}=-1$ so that $\prod_{\hexin \in \Lambda_{\blue{b}}} \hat{B}_{\hex}^{(z)} \prod_{\hexin \in \Lambda_{\red{r}}} \hat{B}_{\hex}^{(z)}=\mathbb{1}$ for periodic boundary conditions respecting the tri-coloring condition.
This results in the following three constraints 
\begin{align}
    \label{Eq::star_constraints}
    \prod_{s\in \Lambda_{\green{g}}} \hat{A}_s = \mathbb{1} \, ,\quad \prod_{s\in \Lambda_{\blue{b}}} \hat{A}_s = \mathbb{1} \, ,\quad \prod_{s\in \Lambda_{\red{r}}} \hat{A}_s = \mathbb{1}\, , 
\end{align}
implying only \(\frac{N}{2}-3\) independent \(\hat{A}_s\) operators.

A similar relation as Eq.~\eqref{Eq::star_constraints} holds for the local symmetries 
\begin{align}
    \prod_{\dodec \in \Lambda_{\green{g}}} \hat{S}_{\dodec}^{(x)} = \mathbb{1} \, ,\quad \prod_{\dodec \in \Lambda_{\blue{b}}} \hat{S}_{\dodec}^{(x)} = \mathbb{1} \, ,\quad \prod_{\dodec \in \Lambda_{\red{r}}} \hat{S}_{\dodec}^{(x)} = \mathbb{1}
\end{align}
in the ground-state manifold, which can again be deduced from $\prod_{\dodec \in \Lambda_{\green{g}}} \hat{S}_{\dodec}^{(x)}\prod_{\hexin \in \Lambda_{\blue{b}}} \hat{B}_{\hex}^{(x)} \prod_{\hexin \in \Lambda_{\red{r}}} \hat{B}_{\hex}^{(x)}=\mathbb{1}$ (and color permutations) and $b_{\hex}^{(x)}=+1$.
Therefore, the number of local symmetry sectors is \(2^{N_{\hex}-3}\).
Comparing the ground-state degeneracy with the number of local symmetry sectors yields $2^6=64$ global sectors, which are of topological origin.
This large number results from additional one-form symmetries besides the \(\hat{\mathcal{W}}^z_n\) and \(\hat{\mathcal{W}}^{x/y}_n\) Wilson loop operators. 
These additional symmetries only commute with the Hamiltonian for \(J_{\tri}=0\), but continue to be relevant as emergent symmetries for small values of \(J_{\tri}\) as discussed below. 

We distinguish three types of one-form symmetries: \(\hat{\mathcal{L}}_{cn}^{x/y} = \prod_{i\in\mathcal{C_{\mathcal{L}}}}\sigma_i^{x/y}\), \(\hat{\mathcal{E}}_{cn}^{x/y} = \prod_{i\in\mathcal{C_{\mathcal{E}}}}\sigma_i^{x/y}\), and \mbox{\(\hat{\mathcal{G}}_{cn}^{z} = \prod_{i\in\mathcal{C_{\mathcal{G}}}}\sigma_i^{z}\)} with eigenvalues $\pm 1$, acting along non-contractible paths \(\mathcal{C}\) on color $c\in\{\green{g},\blue{b},\red{r}\}$ and homology class \(n\in\{1,2\}\) depicted in  Fig.~\ref{Fig::Emergent_symmetries}. Note that each one-form symmetry with color $c$ can only be moved on sublattice $\Lambda_{c}$ through the action of local symmetries.
Not all one-form symmetries are independent, given the constraints
\begin{equation}
\begin{aligned}
    \hat{\mathcal{L}}^{x/y}_{\green{g}n\phantom{b}} \hat{\mathcal{L}}^{x/y}_{\blue{b}n} \hat{\mathcal{L}}^{x/y}_{\red{r}n\phantom{b}} &= \mathbb{1} \\
    \hat{\mathcal{E}}^{x/y}_{\green{g}n\phantom{b}} \hat{\mathcal{E}}^{x/y}_{\blue{b}n} \hat{\mathcal{E}}^{x/y}_{\red{r}n\phantom{b}} &= \hat{\mathcal{W}}_{n\phantom{b}}^{x/y} \\
    \hat{\mathcal{L}}^{x/y}_{cn\phantom{b}} \hat{\mathcal{E}}^{x/y}_{cn\phantom{b}} &= \hat{\mathcal{W}}_{n\phantom{b}}^{x/y} \text{ for } c\in \{\green{g}, \blue{b}, \red{r}\}  \\
    \hat{\mathcal{G}}^{z}_{\green{g}n} \hat{\mathcal{G}}^{z}_{\blue{b}n} \hat{\mathcal{G}}^{z}_{\red{r}n} &= \hat{\mathcal{W}}^{z}_{n}.
\end{aligned}
\end{equation}
Per homology class of the torus, three independent \(\hat{\mathcal{L}}\) or \(\hat{\mathcal{E}}\) symmetries can be chosen, e.g. \(\{\hat{\mathcal{E}}^{x/y}_{\green{g}1},\,\hat{\mathcal{E}}^{x/y}_{\blue{b}1},\,\hat{\mathcal{E}}^{x/y}_{\red{r}1},\,\hat{\mathcal{E}}^{x/y}_{\green{g}2},\,\hat{\mathcal{E}}^{x/y}_{\blue{b}2},\,\hat{\mathcal{E}}^{x/y}_{\red{r}2}\}\). % with $1,2$ denoting the respective homology class. 
No further \(\hat{\mathcal{G}}^{z}_{n}\) operators can be independent, since they do not commute with the selected set. 
Furthermore, either \(x\)- or \(y\)-types of \(\hat{\mathcal{E}}^{x/y}\) may be selected, as the other does not provide further conserved quantities due to 
$\hat{\mathcal{E}}^{y}_{\blue{b}n}\hat{=} \hat{\mathcal{E}}^{x}_{\blue{b}n} \hat{\mathcal{E}}^{z}_{\blue{b}n} = \hat{\mathcal{E}}^{x}_{\blue{b}n} \hat{\mathcal{G}}^{z}_{\red{r}n} \hat{\mathcal{G}}^{z}_{\green{g}n}$,
$\hat{\mathcal{E}}^{y}_{\green{g}n}\hat{=} \hat{\mathcal{E}}^{x}_{\green{g}n} \hat{\mathcal{E}}^{z}_{\green{g}n} = \hat{\mathcal{E}}^{x}_{\green{g}n} \hat{\mathcal{G}}^{z}_{\blue{b}n} \hat{\mathcal{G}}^{z}_{\red{r}n}$,
and $\hat{\mathcal{E}}^{y}_{\red{r}n}\hat{=} \hat{\mathcal{E}}^{x}_{\red{r}n} \hat{\mathcal{E}}^{z}_{\red{r}n} = \hat{\mathcal{E}}^{x}_{\red{r}n} \hat{\mathcal{G}}^{z}_{\green{g}n} \hat{\mathcal{G}}^{z}_{\blue{b}n}$.
In total, six one-form symmetries can be chosen with eigenvalues \(\pm 1\), creating 64 topological sectors.

Next, we discuss the effective low-energy description for small \(J_{\tri}\) resulting from degenerate perturbation theory (see \cite{Suppl} for details of the derivation).
All uneven orders vanish exactly. 
The first non-trivial contribution arises in order six perturbation theory 
\begin{align}
    \label{eq::ham_eff}
    \hat{\mathcal{H}}^{(6)}_{\text{eff},\ZZZ} = - C^{(6)}_{\tri} \sum_{\dodec} \hat{S}_{\dodec}^{(x)}
\end{align}
with \( C^{(6)}_{\tri}= \frac{1}{3456}\frac{J_{\tri}^6}{J_{\hex}^5}\left( \frac{\cos^6{\varphi}}{\sin^5{\varphi}} + \frac{\sin^6{\varphi}}{\cos^5{\varphi}}\right)\).
Keeping in mind that the eigenvalues $s_{\dodec}^{(x)}=\pm 1$ of each $\hat{S}_{\dodec}^{(x)}$ are conserved quantities, one knows exactly the energy spectrum of $ \hat{\mathcal{H}}^{(6)}_{\text{eff}}$.
As $C^{(6)}>0$, the ground-state symmetry sector is fixed and has eigenvalues $s_{\dodec}^{(x)}=+1\,\forall \dodec$.
Any higher-order contribution to the effective model in the infinite system can always be written as a product of $\hat{S}_{\dodec}^{(x)}$ operators so that its spectrum is still diagonal.
This is a direct consequence of the fact that the number of local symmetries matches the Hilbert space dimension of the low-energy subspace.
Consequently, the one-form symmetries \(\hat{\mathcal{L}}_{cn}^{x/y}\), \(\hat{\mathcal{E}}_{cn}^{x/y} \), and \mbox{\(\hat{\mathcal{G}}_{cn}^{z\phantom{/}}\)} are exact symmetries of the effective model to any order, which is not true for the full kagome XYTC, rendering them emergent symmetries.

\begin{figure}[t]
    \centering
    \includegraphics[width=0.7\columnwidth]{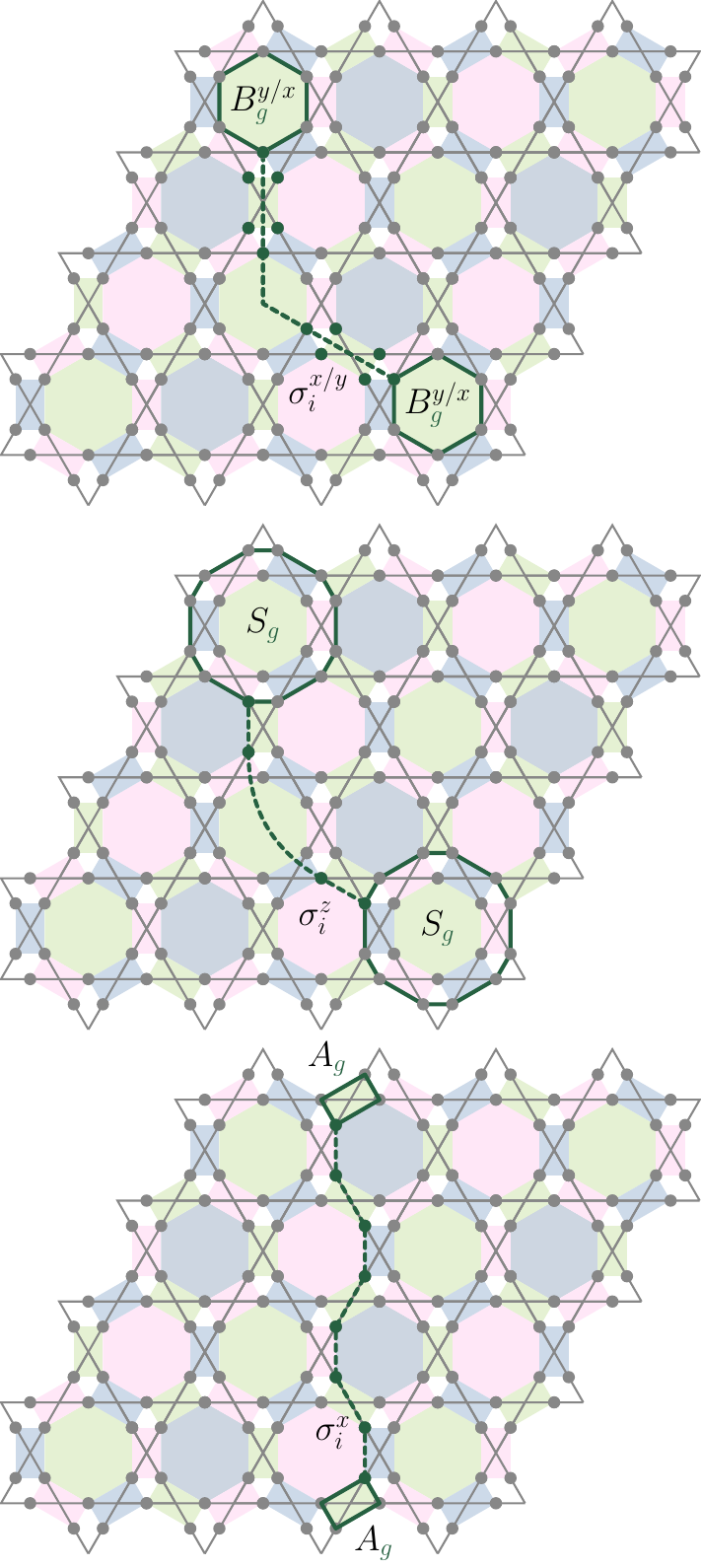}
        \caption{Creation of different elementary particles at the endpoints of strings of Pauli operators. From top to bottom: \(B\) particles, \(S\) particles, and \(A\) particles. The movement of particles is restricted to each sublattice color \(c\in\{\green{g},\blue{b},\red{r}\}\). Here, green is chosen for illustration purposes.}
        \label{Fig::Moving_particles}
\end{figure}

We now consider elementary excitations and mutual particle statistics above the ground state of the effective model \eqref{eq::ham_eff}
with all eigenvalues $s_{\dodec}^{(x)}=+1$ and \mbox{$a_s=b_{\hex}^{(x)}=b_{\hex}^{(y)}=+1$}. Violating either of the ground-state constraints yields elementary excitations referred to as \(B^x\), \(B^y\), \(S\), and \(A\) particles. 
As illustrated in  Fig.~\ref{Fig::Moving_particles}, different particles are created at the endpoints of strings consisting of single-flavor Pauli operators. Per particle type, three colors \(c\in \{\green{g},\blue{b},\red{r}\}\) are distinguished, as each particle can only hop on a sublattice \(\Lambda_{c}\) without creating additional excitations. \(S\) and \(A\) particles are respectively created in pairs on the same sublattice \(\Lambda_{c}\), so that the number of \(S\) and \(A\) particles per sublattice is even.
The string resulting from separating two \(B\) particles on \(\Lambda_{c}\) around the torus is equivalent to \(\hat{\mathcal{L}}\).
Similarly, winding one \(S\) (\(A\)) excitation around the torus corresponds to \(\hat{\mathcal{G}}\) (\(\hat{\mathcal{E}}\)).

This connection to the one-form symmetries suggests non-trivial particle statistics. Indeed, we find the following fusion rules: 
\begin{equation}
\begin{aligned}
    A_{\blue{b}} \times A_{\green{g}} &= B^y_{\red{r}} \\
    B^x_{\blue{b}} \times B^x_{\green{g}} &= B^x_{\red{r}}\\
    B^y_{\blue{b}} \times B^y_{\green{g}} &= B^y_{\red{r}} \\
    S_{\blue{b}} \times S_{\green{g}} &= B^x_{\red{r}}\times B^y_{\red{r}}.
\end{aligned}
\end{equation}
The same holds for any permutation of the three colors.  All elementary \(B\) particles can hence be represented combining \(\{A_{\blue{b}}, A_{\green{g}}, A_{\red{r}}, S_{\blue{b}}, S_{\green{g}}, S_{\red{r}}\}\), yielding \(\mathcal{D}^2=64\) quasiparticles - Abelian anyons with strongly sublattice-dependent mutual statistics. \(S\) particles exhibit trivial statistics with \(A\) particles on different sublattices and semionic statistics on the same sublattice (see supplemental material \cite{Suppl}).
In general, one has \(\gamma = \log \mathcal{D}\), with  \(\gamma\) the topological entanglement entropy and \(\mathcal{D}\) the total quantum dimension \cite{Kitaev_2006_b,Levin_2006}. \(\mathcal{D}^2\) matches the number of topological superselection sectors, here established by the six independent emergent one-form symmetries, and we accordingly find 
\(\mathcal{D}=8\).
For small \(J_{\tri}\), the kagome XYTC hence displays \(\Z _2 \times \Z _2 \times \Z _2\) topological order.

%Z2 topological order
%%%%%%%%%%%%%%%%%%%%%%%%%%%%%%%%%%%%%%%%%%%%%%%%%%%%%%%%%%%%%%%%%%%%%%%%%%%%%%%%%%%%%%%%%%%%
{\it{{\it $\ZZ$ topological order}:}}
Aside from this \(\ZZZ\) phase, we find that the remainder of the phase diagram exhibits \(\Z_2\) topological order.
The system reduces to a conventional toric code on the kagome lattice in the cases \(\varphi = 0\) and \(\varphi = \frac{\pi}{2}\) (and all \(J_{\tri}\)). In this instance, all plaquette operators commute with the Hamiltonian, making both stars and plaquettes conserved quantities. Due to topological order, the phase is stable for small deviations of \(\varphi\) from these limits. Importantly, \mbox{\(s_{\dodec}^{(x)} = s_{\dodec}^{(y)} = +1\)} in the ground-state manifold. Whenever the kagome XYTC is adiabatically connected to the toric code limits, it exhibits \(\Z_2\) topological order.

Let us now address the limit \mbox{\(J_{\hex} = 0\)} of isolated triangular plaquettes, where four local ground states with \(b_{\tri}=+1\) are present per triangle for all values of \(\varphi\). We stress that this includes \(\varphi = 0\) and \(\varphi = \frac{\pi}{2}\). The ground-state manifold with \(a_s = +1\,\forall s\) is extensively degenerate (\(\#\text{GS} = 2^{N_{\hex}+1}\)) when periodic boundary conditions are imposed. In contrast to the previous limit \mbox{\(J_{\tri} = 0\)}, star operators are solely constrained by \(\prod_s \hat{A}_s = \mathbb{1}\). Four topological sectors corresponding to the exact \(\hat{\mathcal{W}}_{n}^{\alpha}\) symmetries are established, each  \(2^{N_{\hex}-1}\)-fold degenerate due to the presence of the local \(\hat{S}_{\dodec}^{(x)}\) symmetries, satisfying \(\prod_{\dodec} \hat{S}_{\dodec}^{(x)} = \mathbb{1}\). Crucially, for each topological sector and fixed \(s_{\dodec}^{(x)}\) eigenvalues, there is one unique ground state.

First-order degenerate perturbation theory in this ground-state manifold yields, up to a constant, the exactly solvable effective low-energy Hamiltonian
\begin{align}
    \label{eq::ham_eff_Z2}
    \hat{\mathcal{H}}^{(1)}_{\text{eff},\ZZ} = - C_{\hex}^{(1)} \sum_{\dodec} \hat{S}_{\dodec}^{(x)}
\end{align}
with \(C_{\hex}^{(1)} = J_{\hex} \left(\cos^7{\varphi} + \sin^7{\varphi} \right) \)
for small \(J_{\hex}>0\), as derived in the supplemental material \cite{Suppl}. The ground states are fixed by the eigenvalues \(s_{\dodec}^{(x)}=+1\,\forall \dodec\) and therefore completely determined by the eigenvalues of the exact Wilson loop operators \(\hat{\mathcal{W}}^{\alpha}_{n}\). Hence, a \(\Z _2\) topologically ordered phase is established for all \(\varphi\) and small \( J_{\hex}\) in the thermodynamic limit.

%Figure 3 - Moving particles
%%%%%%%%%%%%%%%%%%%%%%%%%%%%%%%%%%%%%%%%%%%%%%%%%%%%%%%%%%%%%%%%%%%%%%%%%%%%%%%%%%%%%%%%%%%%
%%%%%%%%%%%%%%%%%%%%%%%%%%%%%%%%%%%%%%%%%%%%%%%%%%%%%%%%%%%%%%%%%%%%%%%%%%%%%%%%%%%%%%%%%%%%

\begin{figure}[t]
       \centering
        \includegraphics[width=0.9\columnwidth]{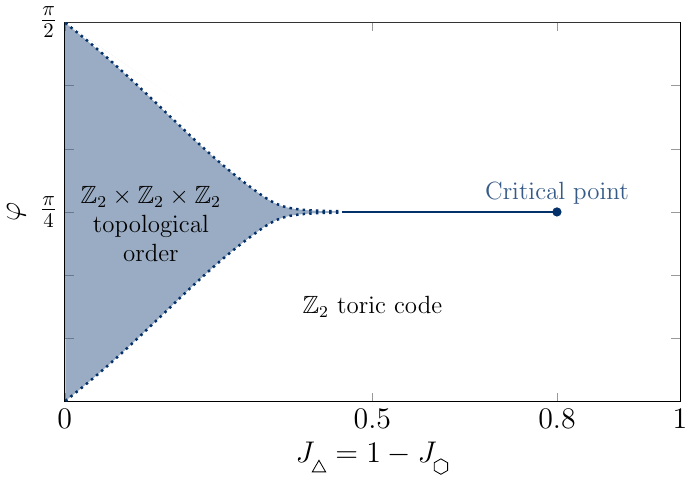}
        \caption[Quantum phase diagram of the kagome XYTC]{
        Quantum phase diagram of the kagome XYTC, hosting \(\ZZZ\) and a \(\Z_2\) topologically ordered phases depending on the anisotropy of plaquettes $J_{\tri}$ and the XY-anisotropy \(\varphi\). 
        The phase diagram is symmetric about reflections along \(\varphi = \frac{\pi}{4}\). All lines correspond to first-order transitions.
        The solid (dotted) line has been determined by series expansions (exact diagonalization).
        A critical point at \(J_{\tri} \approx 0.8\) occurs at the end of the \(\varphi = \frac{\pi}{4}\)-phase transition line. 
        }
       \label{Fig:XYTC_pd}
\end{figure}

%Quantum Phase Diagram
%%%%%%%%%%%%%%%%%%%%%%%%%%%%%%%%%%%%%%%%%%%%%%%%%%%%%%%%%%%%%%%%%%%%%%%%%%%%%%%%%%%%%%%%%%%%
{\it{{\it Quantum phase diagram}:}}
So far, we have demonstrated that the kagome XYTC realizes two distinct phases with topological order - \(\Z_2\) topological order in the case of small \(J_{\hex}\) and \(\ZZZ\) topological order for small \(J_{\tri}\).
In Fig.~\ref{Fig:XYTC_pd}, the full quantum phase diagram is shown as a function of the anisotropy of plaquettes \(J_{\tri}\) and the XY-anisotropy \(\varphi\).
The phase diagram is symmetric under \(\varphi \rightarrow \frac{\pi}{2}-\varphi\) due to \( \hat{\mathcal{U}}^{\dagger}\hat{\mathcal{H}}(\varphi) \hat{\mathcal{U}}^{\phantom{\dagger}}  = \hat{\mathcal{H}}(\frac{\pi}{2}-\varphi)\) with \mbox{\( \hat{\mathcal{U}}= \exp{( -\mathrm{i}\frac{\pi}{4} \sum_j \sigma^{z}_j)}\)}.

The quantum phase diagram has been deduced from high-order series expansions and exact diagonalization. For details, we refer to the supplemental material \cite{Suppl}. Phase transitions between both \(\Z_2\) phases are located from order-14 series expansions of the ground-state energy per site about the exactly solvable toric code limits in the parameter \(\tan\varphi\), depending on the ratio \(\frac{J_{\tri}}{J_{\hex}}\). We find first-order phase transitions at the self-dual line \mbox{\(\varphi = \frac{\pi}{4}\)} for intermediate values of \(J_{\tri}\) (see solid line in Fig.~\ref{Fig:XYTC_pd}). The line terminates at a critical end point for \(J_{\tri}\approx 0.8\). 
Phase transitions between \(\ZZZ\) and \(\Z_2\) phases are determined from exact diagonalization for a system comprising 24 spins, and are also first-order (see dotted line in Fig.~\ref{Fig:XYTC_pd}) so that finite-size effects are expected to be small. The determined phase transition line is qualitatively consistent with the one obtained from comparing the order-14 series of the ground-state energy of the \(\Z_2\) phases with the order-6 one of the \(\ZZZ\) phase about the limits \(\varphi = 0\) and \(J_{\tri}=0\) \cite{Suppl}. 
 
%Figure 4 - Quantum phase diagram kagome XYTC
%%%%%%%%%%%%%%%%%%%%%%%%%%%%%%%%%%%%%%%%%%%%%%%%%%%%%%%%%%%%%%%%%%%%%%%%%%%%%%%%%%%%%%%%%%%%
%%%%%%%%%%%%%%%%%%%%%%%%%%%%%%%%%%%%%%%%%%%%%%%%%%%%%%%%%%%%%%%%%%%%%%%%%%%%%%%%%%%%%%%%%%%%

%Conclusions
%%%%%%%%%%%%%%%%%%%%%%%%%%%%%%%%%%%%%%%%%%%%%%%%%%%%%%%%%%%%%%%%%%%%%%%%%%%%%%%%%%%%%%%%%%%%
{\it{Conclusions:}}
In this work we have determined the quantum phase diagram of the kagome XYTC based on analytical considerations, series expansions, and exact diagonalizations. We have shown analytically that an unconventional \(\ZZZ\) topological order with \(\mathcal{D}=8\) is present for small \(J_{\tri}\), hosting 64 Abelian anyons as quasiparticles. Additional emergent one-form symmetries were discovered, explaining the large number of topological superselection sectors. The remainder of the phase diagram exhibits \(\Z_2\) topological order. All phase transitions are first order except at a critical end point.

Our work demonstrates that geometry strongly influences the quantum phases of the XYTC. While the model displays a subsystem-symmetry-broken phase on the square lattice \cite{Vieweg_2025,Qiao_2025}, the kagome XYTC is topologically ordered in the whole parameter space except \(J_{\tri}\in\{0,1\}\). It would be interesting to generalize these results to other lattice geometries and topological phases. For both, the square and kagome XYTC, the mobility of quasiparticles is restricted. On the square lattice, the quasiparticles behave as fractons and lineons, whereas on the kagome lattice in the \(\ZZZ\) phase, the particles are restricted to colored sublattices. This may indicate a similar origin of these phases as condensates of extended objects resembling 3D fracton models \cite{Ma_2017}.

The \(\ZZZ\) topological order of the kagome XYTC is similar to the \(\Z _2 \times \Z _2\) topological phase of the hexagonal color code (CC) in that three colors play a crucial role \cite{Bombin_2006}. The kagome XYTC enhances the \(\Z _2 \times \Z _2\) phase of the CC through two additional independent one-form symmetries.
The Abelian phase is therefore richer than that of the CC and the conventional toric code.
Moreover, the effective model \eqref{eq::ham_eff} for small \(J_{\tri}\) can be transformed to a CSS stabilizer code \cite{Calderbank_1996,Steane_1996}.
CSS codes play a central role in current approaches for quantum error correction.
To this end, we consider the topological stabilizer code (SC) 
\begin{align}
    \hat{\mathcal{H}}_{\mathrm{SC}} = -\sum_s \hat{A}_s - \sum_{\hex} \hat{B}_{\hex}^{(y)} - \sum_{\hex} \hat{B}_{\hex}^{(x)} - \sum_{\dodec} \hat{S}_{\dodec}^{(x)}\,,
\end{align}
which reduces exactly to $\eqref{eq::ham_eff}$ in the manifold given by $a_s=b^{(x)}_{\hex}=b^{(y)}_{\hex}=+1$ on all stars and hexagons. 
This SC is not a CSS code due to the presence of all Pauli flavors. 
However, when replacing \(\hat{B}^{(y)}_{\hex}\) by \(\hat{B}^{(z)}_{\hex}\), one gets a distinct CSS code with topological ground-state degeneracy \mbox{\(\mathcal{D}^2=64\)} on the torus. This is very promising for the encoding and manipulation of topological quantum information.

%Acknowledgments
%%%%%%%%%%%%%%%%%%%%%%%%%%%%%%%%%%%%%%%%%%%%%%%%%%%%%%%%%%%%%%%%%%%%%%%%%%%%%%%%%%%%%%%%%%%%
{\it{Acknowledgments:}}
We thank Harald Leiser, Andreas Schellenberger, and Viktor Kott for fruitful discussions. KPS acknowledges financial support by the German Science Foundation (DFG) through the Munich Quantum Valley, which is supported by the Bavarian state government with funds from the Hightech Agenda Bayern Plus.

\paragraph{Author contributions}
C.K. performed the series expansions and the numerical calculations.
C.K. and M.V. worked out the analytical arguments and analyzed the numerical results.
M.V. and K.P.S. conceptualized the work.
All authors contributed to the writing of the draft.

\bibliography{bibliography_2.bib}
%================================================================================================================================================================================================================================================================================================================================
\newpage
%Nummerierung beginnt wieder bei 1
\setcounter{equation}{0}
\setcounter{figure}{0}
%Hyperref-Anker neu
\renewcommand{\theHequation}{II.\arabic{equation}}
\renewcommand{\theHfigure}{II.\arabic{figure}}

%%%%%%%%%
\onecolumngrid
\begin{center}
{\large\bfseries Supplemental Material for \enquote{Unconventional $\ZZZ$ topological order in the kagome XY toric code}}
\end{center}
\twocolumngrid
This Supplemental Material contains additional information regarding the series expansions for small \(J_{\tri}\) and \(J_{\hex}\), details about the ground-state manifold and quasiparticle statistics of the \(\ZZZ\) topologically ordered phase, and a deduction of the quantum phase diagram based on series expansions and numerical exact diagonalization.

\section[Degenerate perturbation theory for both J-limits]{Degenerate perturbation theory for small \texorpdfstring{$J_{\trirobust}$ and $J_{\hexrobust}$}{Degenerate perturbation theory for both J-limits}}

\subsection[Perturbation theory for small J\textsubscript{tri}]{Perturbation theory for small \texorpdfstring{$J_{\trirobust}$}{Perturbation theory for small  J\textsubscript{tri}}} \label{sec:Perturbation_theory_small_triangles}

We derive the sixth-order effective low-energy Hamiltonian for small \(J_{\tri}\) from Takahashi's degenerate perturbation theory \cite{Takahashi_1977, Mila_2010} based on the extensively degenerate unperturbed ground-state manifold \mbox{(\(J_{\tri}=0\))}. The problem can be stated in the form \mbox{\(\hat{\mathcal{H}} = \hat{\mathcal{H}}_0 + J_{\tri} \hat{V}\)} with perturbation parameter \(J_{\tri}\), perturbation \(\hat{V}=-\sum_{\tri} \hat{B}_{\tri}\), and unperturbed Hamiltonian \mbox{\(\hat{\mathcal{H}}_0=-J_s\sum_s \hat{A}_s - J_{\hex}\sum_{\hex}\hat{B}_{\hex}\)}. Local eigenstates of \(\hat{B}_{\hex}\) are linear combinations \mbox{\(\ket{\hexstatePlatzhalter\,}\pm B_{\hex}^{(x)} \ket{\hexstatePlatzhalter\,}\)} with an arbitrary number of up and down spins. Depending on the four different eigenvalues \mbox{\(\cos{\varphi}+\sin{\varphi}\)}, \mbox{\(-\cos{\varphi}+\sin{\varphi}\)}, \mbox{\(+\cos{\varphi}-\sin{\varphi}\)}, and \mbox{\(-\cos{\varphi}-\sin{\varphi}\)}, we denominate the eigenstates \(\ket{\mathrm{GS}}_l\), \(\ket{\mathrm{Ex}^{(x)}}_l\), \(\ket{\mathrm{Ex}^{(y)}}_l\), and \(\ket{\mathrm{Ex}^{(xy)}}_l\) in respective order with \(l\in\{1,...,16\}\):
\begin{equation}
\begin{aligned}
    \ket{\mathrm{GS}}_l &= \frac{\ket{\hexstateuneven}_l + \hat{B}_{\hex}^{(x)} \ket{\hexstateuneven}_l}{\sqrt{2}} \\
    \ket{\mathrm{Ex}^{(x)}}_l &= \frac{\ket{\hexstateeven}_l - \hat{B}_{\hex}^{(x)} \ket{\hexstateeven}_l}{\sqrt{2}} \\
    \ket{\mathrm{Ex}^{(y)}}_l &= \frac{\ket{\hexstateeven}_l + \hat{B}_{\hex}^{(x)} \ket{\hexstateeven}_l}{\sqrt{2}} \\
    \ket{\mathrm{Ex}^{(xy)}}_l &= \frac{\ket{\hexstateuneven}_l - \hat{B}_{\hex}^{(x)} \ket{\hexstateuneven}_l}{\sqrt{2}}\, ,
\end{aligned}
\end{equation}
%\begin{equation}
%\begin{aligned}
%    \ket{\mathrm{GS}}_l &= \ket{\hexstateuneven}_l &&+ \hat{B}_{\hex}^{(x)} \ket{\hexstateuneven}_l \\
%    \ket{\mathrm{Ex}^{(x)}}_l &= \ket{\hexstateeven}_l &&- \hat{B}_{\hex}^{(x)} \ket{\hexstateeven}_l \\
%    \ket{\mathrm{Ex}^{(y)}}_l &= \ket{\hexstateeven}_l &&+ \hat{B}_{\hex}^{(x)} \ket{\hexstateeven}_l \\
%    \ket{\mathrm{Ex}^{(xy)}}_l &= \ket{\hexstateuneven}_l &&- \hat{B}_{\hex}^{(x)} \ket{\hexstateuneven}_l\, ,
%\end{aligned}
%\end{equation}
where even (uneven) denotes the parity of up spins on a given hexagon. Acting with \(\sigma_j^x\) and \(\sigma_{k}^y\) on the ground states results in
%%%%%%%
\begin{equation}
\begin{alignedat}{2}
    \sigma_j^x \ket{\mathrm{GS}}_l &= \ket{\mathrm{Ex}^{(y)}}_m \\
    \sigma_k^y \ket{\mathrm{GS}}_l &=  \pm \mathrm{i} \ket{\mathrm{Ex}^{(x)}}_m \\
    \sigma_j^x \sigma_k^y \ket{\mathrm{GS}}_l &= \pm \mathrm{i} \ket{\mathrm{Ex}^{(xy)}}_m 
\end{alignedat}
\end{equation}
with arbitrary \(j,k\in\ininhex\), \(k\)-dependent factor \(\pm \mathrm{i}\), and \(j,k,l\)-dependent index \(m\in\{1,...,16\}\).

\begin{figure}[t]
    \centering
    \includegraphics[width=0.4\linewidth]{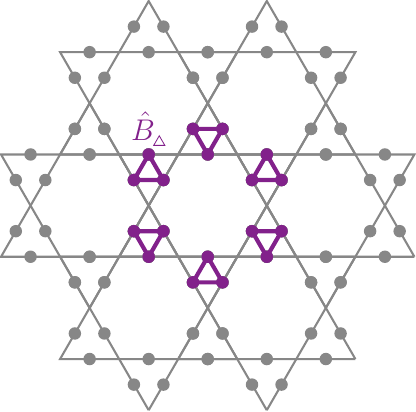}
    \caption[The only non-trivial contribution in order-6 perturbation theory. Six \(\hat{B}_{tri}\) operators depicted in purple act along a hexagon.]{The only non-trivial contribution in order-6 perturbation theory. Six triangular \(\hat{B}_{\tri}\) operators depicted in purple act along a hexagon.}
    \label{fig:Perturbation_O6}
\end{figure}

In Takahashi's degenerate perturbation theory, the effective Hamiltonian is expressed as operator sequences combining \(\hat{P}_0\), \(\hat{V}\), and \(\hat{S}\) \cite{Takahashi_1977,Mila_2010}, where \(\hat{P}_0\) is a projector onto the unperturbed eigenspace \(U_0\) and \(\hat{S} = \frac{1-\hat{P}_0}{E_0-\hat{\mathcal{H}}_0}\) a weighted projector onto its orthogonal subspace. In the given problem, only by acting an even number of times with perturbation \(\hat{V}\) can a hexagon return to a ground state. Therefore, all uneven orders vanish necessarily. Second and fourth orders produce only constant diagonal contributions, as each hexagon must be perturbed twice by the same operator to return to a ground state. In order six, the only non-diagonal term appears as a result of the operator sequence \(\hat{P}_0 \hat{V} \hat{S} \hat{V} \hat{S} \hat{V} \hat{S} \hat{V} \hat{S} \hat{V} \hat{S} \hat{V} \hat{P}_0\) acting along the configuration depicted in Fig.~\ref{fig:Perturbation_O6}.  
The only non-vanishing contributions corresponding to this operator sequence are processes where only \(\hat{B}_{\tri}^{(x)}\) or only \(\hat{B}_{\tri}^{(y)}\) operators act along the configuration \(\sixtriangles\,\) shown in Fig.~\ref{fig:Perturbation_O6}. All other combinations are projected to zero.
To calculate the corresponding matrix elements, first, the energy term \(1/(E_0 - \hat{\mathcal{H}}_0 )\) has to be evaluated for each state with a single type of excitation.
For a product state $\ket{\psi_n}^{(x)}$ ($\ket{\psi_n}^{(y)}$) with $n$ local excitations \(\ket{\mathrm{Ex}^{(x)}}_l\) (\(\ket{\mathrm{Ex}^{(y)}}_l\)), we obtain the expressions
\begin{equation}
\begin{aligned}
\bra{\psi_n}^{(x)}\frac{1}{E_0 - \hat{\mathcal{H}}_0}\ket{\psi_n}^{(x)}&= \frac{1}{-2nJ_{\hex}\cos{\varphi}} \equiv \frac{1}{x_n} \\
\bra{\psi_n}^{(y)}\frac{1}{E_0 - \hat{\mathcal{H}}_0}\ket{\psi_n}^{(y)}&= \frac{1}{-2n J_{\hex}\sin{\varphi}} \equiv \frac{1}{y_{n}}.
\end{aligned}
\end{equation}
For the process involving \(\hat{B}_{\tri}^{(x)}\), we thus compute the matrix element \mbox{\(C_{\tri}^{(6,x)} \equiv \frac{\cos^6{\varphi}}{y_3^3y_2^2}=\frac{1}{J_{\hex}^5}\frac{1}{3456}\frac{\cos^6{\varphi}}{\sin^5{\varphi}}\)}, and for the process involving \(\hat{B}_{\tri}^{(y)}\), \mbox{\(C_{\tri}^{(6,y)} \equiv \frac{\sin^6{\varphi}}{x_3^3x_2^2}=\frac{1}{J_{\hex}^5}\frac{1}{3456}\frac{\sin^6{\varphi}}{\cos^5{\varphi}}\)}. 
In total, we find
\begin{equation}
\begin{aligned}
J_{\tri}^6 C_{\tri}^{(6,x)} \prod_{\tri \in \sixtriangles\,}\hat{B}_{\tri}^{(x)}+ J_{\tri}^6 C_{\tri}^{(6,y)}\prod_{\tri \in \sixtriangles\,}\hat{B}_{\tri}^{(y)} \\
= J_{\tri}^6 C_{\tri}^{(6,x)} \hat{S}_{\dodec}^{(x)}\hat{B}_{\hex}^{(x)} + J_{\tri}^6 C_{\tri}^{(6,y)}\hat{S}_{\dodec}^{(y)}\hat{B}_{\hex}^{(y)} 
\end{aligned}
\end{equation}
for each \(\sixtriangles\,\). Since \(\hat{S}_{\dodec}^{(y)}=\hat{S}_{\dodec}^{(x)}\) and \(b_{\hex}^{(x)}= b_{\hex}^{(y)}=1\) in the ground-state manifold, the effective Hamiltonian in order six perturbation theory is given by
\begin{align}
    \label{eq::ham_eff_supp}
    \hat{\mathcal{H}}^{(6)}_{\text{eff},\ZZZ} = E^{(6)}_0  - C^{(6)}_{\tri} \sum_{\dodec} \hat{S}_{\dodec}^{(x)}
\end{align}
with \( C^{(6)}_{\tri}= \frac{1}{3456}\frac{J_{\tri}^6}{J_{\hex}^5}\left( \frac{\cos^6{\varphi}}{\sin^5{\varphi}} + \frac{\sin^6{\varphi}}{\cos^5{\varphi}}\right)\).
Additionally, the constant contribution is calculated to be
\begin{equation} \label{eq:Effective_O6_Hamiltonian}
\begin{aligned}
\frac{E_0^{(6)}}{N} =  
    &- \frac{J_{\tri}^2}{J_{\hex}}\frac{1}{18}\left(\frac{\sin^2{\varphi}}{\cos{\varphi}}+\frac{\cos^2{\varphi}}{\sin{\varphi}}\right) \\
    &-\frac{J_{\tri}^4}{J_{\hex}^3}\frac{19}{1296}\left(\frac{\sin^4{\varphi}}{\cos^3{\varphi}}+\frac{\cos^4{\varphi}}{\sin^3{\varphi}}\right) \\
    &-\frac{J_{\tri}^6}{J_{\hex}^5}\frac{171173}{6531840}\left(\frac{\sin^6{\varphi}}{\cos^5{\varphi}}+\frac{\cos^6{\varphi}}{\sin^5{\varphi}}\right)\,.
\end{aligned}
\end{equation}

%%%%%%%%%%%%
\subsection[Perturbation theory for small J\textsubscript{hex}]{Perturbation theory for small \texorpdfstring{$J_{\hexrobust}$}{Perturbation theory for small  J\textsubscript{hex}}}

In the opposite limit for small \(J_{\hex}\), we perform Takahashi's degenerate perturbation theory to derive the exact first-order effective Hamiltonian. The problem reads \mbox{\(\hat{\mathcal{H}} = \hat{\mathcal{H}}_0 + J_{\hex} \hat{V}\)} with perturbation parameter \(J_{\hex}\), perturbation \(\hat{V}=-\sum_{\hex} \hat{B}_{\hex}\), and unperturbed Hamiltonian \mbox{\(\hat{\mathcal{H}}_0=-J_s\sum_s \hat{A}_s - J_{\tri}\sum_{\tri}\hat{B}_{\tri}\)}. As \(\hat{B}_{\tri}\) is unitary and hermitian, the eigenvalues are given by \(b_{\tri}=\pm 1\). The corresponding four local ground states on each triangle are
\begin{equation}
\begin{aligned}
\ket{\psi_1} &= \frac{1}{\sqrt{2}}\left(e^{\mathrm{i}\varphi} \ket{\up\up\up}+\ket{\down\down\down}\right) 
\\
\ket{\psi_2} &= \frac{1}{\sqrt{2}}\left(e^{-\mathrm{i}\varphi} \ket{\up\up\down}+\ket{\down\down\up}\right) 
\\
\ket{\psi_3} &= \frac{1}{\sqrt{2}}\left(e^{-\mathrm{i}\varphi} \ket{\down\up\up}+\ket{\up\down\down}\right) 
\\
\ket{\psi_4} &= \frac{1}{\sqrt{2}}\left(e^{-\mathrm{i}\varphi} \ket{\up\down\up}+\ket{\down\up\down}\right).
\end{aligned}
\end{equation}
Excited states corresponding to the eigenvalue \(b_{\tri}=-1\) differ by a minus sign in front of the second summand.

We now analyze the first-order term 
\begin{equation}
\hat{P}_0 \hat{V} \hat{P}_0= -\sum_{\hex} \hat{P}_0 \left(\cos{\varphi}\hat{B}_{\hex}^{(x)}+\sin{\varphi}\hat{B}_{\hex}^{(y)} \right) \hat{P}_0
\end{equation}
more closely. 
In fact, the relation
\begin{align}
    \hat{B}_{\hex}^{(\alpha)} = \prod_{\tri \in \sixtriangles\,} \hat{B}_{\tri}^{(\alpha)} \hat{S}_{\dodec}^{(\alpha)}
\end{align}
holds, where each \(\hat{B}_{\tri}^{(\alpha)}\) operator affects one local triangle. Hence, the matrix elements can easily be calculated to be 
\begin{equation}
\begin{aligned}
    \bra{\psi}\hat{P}_0 \prod_{\tri \in \sixtriangles\,} \hat{B}_{\tri}^{(x)}\hat{P}_0 \ket{\psi} &= \cos^6{\varphi} \\
    \bra{\psi}\hat{P}_0 \prod_{\tri \in \sixtriangles\,} \hat{B}_{\tri}^{(y)}\hat{P}_0 \ket{\psi} &= \sin^6{\varphi},
\end{aligned}
\end{equation}
where $\ket{\psi}$ is an element of the ground-state manifold. 
In total, up to a trivial constant in order-zero perturbation theory, we find the first-order effective Hamiltonian
\begin{align}
    \hat{\mathcal{H}}^{(1)}_{\text{eff},\Z_2} = - C_{\hex}^{(1)} \sum_{\dodec} \hat{S}_{\dodec}^{(x)}
\end{align}
with \(C_{\hex}^{(1)} = J_{\hex} \left(\cos^7{\varphi} + \sin^7{\varphi} \right) \)
for small \(J_{\hex}>0\). 

%%%%%%%%%%%%%
\section[Details about the ZZZ topologically ordered phase]{Details about the \(\ZZZ\) topologically ordered phase}
\subsection{Non-emptiness of the ground-state sector}
The ground-state sector of the \(\ZZZ\) topologically ordered phase characterized by \mbox{\(b_{\hex}^{(x)}=b_{\hex}^{(y)}=a_s=s_{\dodec}^{(x)}=1\)} is non-empty, as can be seen from the following argument. We start with the non-zero state \mbox{\(\ket{\psi}=\prod_{\hex} \left( \mathbb{1} + \hat{B}_{\hex}^{(z)} \right) \prod_{s} \left( \mathbb{1} + \hat{A}_s \right) \ket{\Rightarrow}\)}.
Here, $\ket{\Rightarrow}$ denotes the product state where all spins point in the $x$-direction. This state is non-zero since expanding the product gives an equal-weight superposition of orthonormal product states with only positive coefficients. The state $\ket{\psi}$ trivially satisfies \mbox{\(b_{\hex}^{(x)}=s_{\dodec}^{(x)}=1\)}.
Based on this configuration, a ground state can be constructed by flipping all \(\hat{B}_{\hex}^{(z)}\) eigenvalues from \(+1\) to \(-1\) through the action of 
$\hat{B}_{\tri}^{(x)}$ operators. 
This is always possible for a system with periodic boundary conditions that respects the tricolor division.
Indeed, the procedure yields a ground state due to \mbox{\(\left( \mathbb{1} + \hat{B}_{\hex}^{(y)} \right) \ket{\Rightarrow} = \left( \mathbb{1} - \hat{B}_{\hex}^{(z)} \right) \ket{\Rightarrow}\)}. 
%\(x\)-flavored \(\twotrianglesonerectangle\)-operators connecting each pair of same-colored hexagons. Each \(\twotrianglesonerectangle\)-operator commutes with all \(\hat{A}_s\) and other \(\hat{B}_{\hex}^{(z)}\) operators. This works for all systems with reasonably imposed periodic boundary conditions respecting the tricolor division. The procedure indeed yields a ground state due to \mbox{\(\left( \mathbb{1} + \hat{B}_{\hex}^{(y)} \right) \ket{\Rightarrow} = \left( \mathbb{1} - \hat{B}_{\hex}^{(z)} \right) \ket{\Rightarrow}\)}. 

%%%%%%%%%%%%%
\subsection{Demonstration of particle statistics}
\begin{figure}[h]
    \centering
    \includegraphics[width=0.9\linewidth]{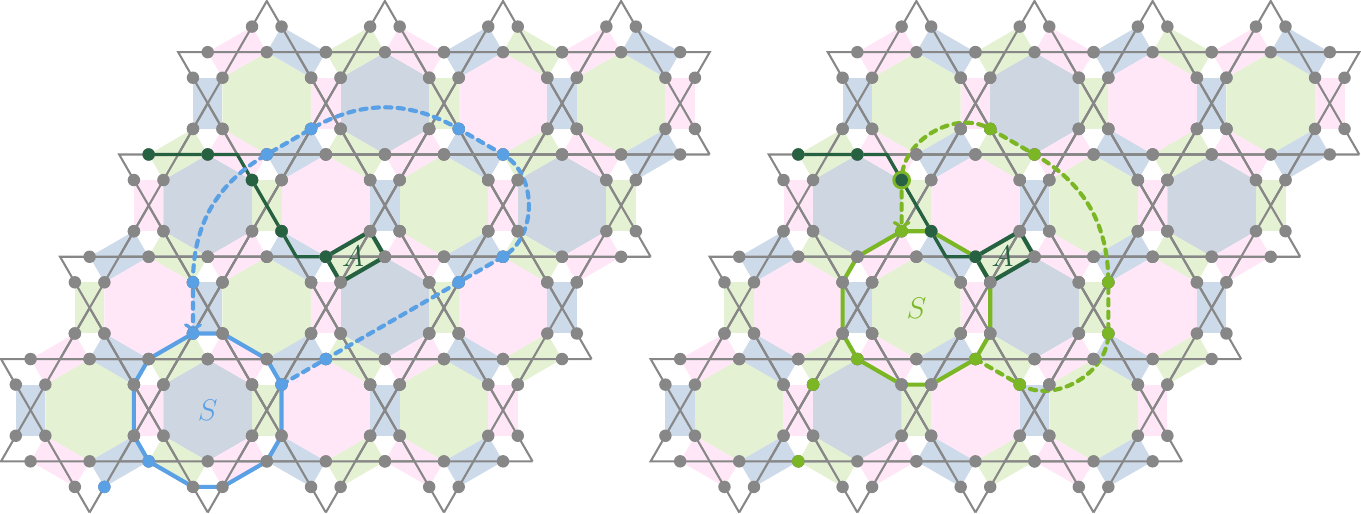}
    \caption{Winding an \(S\) particle around an \(A\) particle. \textit{Left}: Mutual statistics on differently colored sublattices (here blue and green). \textit{Right}: Mutual statistics on the same sublattice (here green). The winding path of the \(S\) particle crosses the construction path of the \(A\) particle in a single spin, illustrated by the dark green and light green filled circle in the upper left.}
    \label{fig:Mutual_particle_statistics}
\end{figure}

Fig.~\ref{fig:Mutual_particle_statistics} illustrates the winding process of one \(S\) particle around an \(A\) particle. Recall that \(S\) (\(A\)) particles are created and moved through a string of Pauli \(z\) (\(x\)) operators. If both particles are differently colored, the mutual statistics are trivial due to no coinciding spins involved in the winding process. However, if the particles reside on the same sublattice, moving the \(S\) particle around the \(A\) particle, their paths cross in one single spin, resulting in an overall minus sign of the wave function.
%
\iffalse
Suppose \(S_{\green{g}}\) (\(A_{\green{g}}\)) is an endpoint of the string operator \(L^z\) (\(L^x\)). The initial state hence is \mbox{\(\ket{\psi_\text{initial}}=L^zL^x\ket{\phi}\)} with \(\ket{\phi}\) a ground state and \([L^z,\,L^x]=0\) per construction. \(S_{\green{g}}\) is moved around \(A_{\green{g}}\) through the action of \(W^z\). Thus, the final state is
\begin{equation}
\begin{aligned}
    \ket{\psi_{\text{final}}}&=W^z\ket{\psi_\text{initial}} =W^zL^zL^x\ket{\phi} \\&=-L^xL^zW^z\ket{\phi} = -L^zL^x\ket{\phi} \\ &=-\ket{\psi_\text{initial}},
\end{aligned}
\end{equation}
because \(L^x\) and \(W^z\) anticommute.
\fi

%%
\section{Deduction of the quantum phase diagram}
The quantum phase diagram is deduced from high-order series expansions and numerical exact diagonalization. Due to naturally incorporating \(a_s=1\,\forall s\) constraints, the series expansions of the ground-state energy per site based on \(\varphi = 0\) are calculated in a dual model of the kagome XYTC,  which is introduced in sec.~\ref{sec:Dual_model}. Exact diagonalization is described in sec.~\ref{sec:ED}.
\subsection{Dual model}\label{sec:Dual_model}
\begin{figure}[t]
    \centering
    \includegraphics[width=0.9\linewidth]{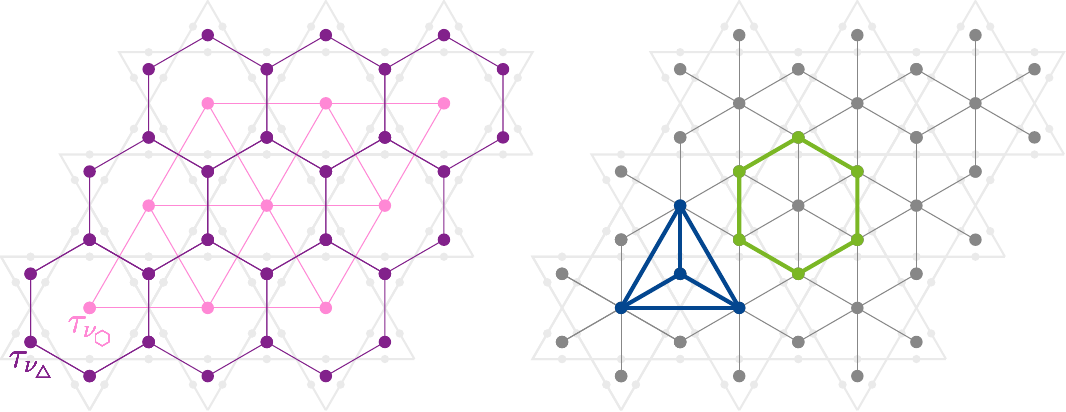}
    \caption[Placing pseudospins at the center of triangular (hexagonal) plaquettes yields a hexagonal (triangular) sublattice depicted in purple (pink). \textit{Right}: Four- (blue) and six-spin (green) interactions in the dual model of the kagome XYTC on the dice lattice.]{\textit{Left}: Placing pseudospins \(\tau_{\nu_{\tri}}\) (\(\tau_{\nu_{\hex}}\)) at the center of triangular (hexagonal) plaquettes yields a hexagonal (triangular) sublattice depicted in purple (pink). \textit{Right}: Four- (blue) and six-spin (green) interactions in a dual model of the kagome XYTC on the dice lattice.}
    \label{fig:Dual}
\end{figure}

A dual model of the kagome XYTC can be derived from an isospectral duality mapping \cite{Kramers_1941, Savit_1980} in the low-energy sector characterized by \(a_s = 1\,\forall s\). Pseudospins \(1/2\) are introduced in the center of all plaquettes so that eigenvalues $\pm 1$ of $\hat{B}_{\tri/\hex}^{(x)}$ operators correspond to diagonal entries of the Pauli \(z\)-matrix. 
Pseudospins $\tau^{\alpha}_{\nu_{\hex}}$ placed in the center of hexagonal plaquettes $\nu_{\hex}$ form a triangular lattice (Tri), while pseudospins $\tau^{\alpha}_{\nu_{\tri}}$ situated in the middle of triangular plaquettes $\nu_{\tri}$ build a hexagonal sublattice (Hex), as depicted in Fig.~\ref{fig:Dual}. Together, a dice lattice is established. 
The mapping based on \(\varphi = 0\) then yields the Hamiltonian
\begin{equation} \label{eq:Hamiltonian_dual_model}
\begin{aligned}
     \hat{\mathcal{H}}_{\mathrm{dual}} &=  - J_{\tri} \cos{\varphi} \sum_{\mathrm{Hex}}\tau^z_{\nu_{\tri}}   - J_{\hex} \cos{\varphi} \sum_{\mathrm{Tri}}\tau^z_{\nu_{\hex}} \\ &\quad - J_{\tri} \sin{\varphi} \sum_{\centertriangle} \tau_{\nu_{\tri}}^x \prod_{\mathclap{\nu_{\hex} \in \trianglewithoutcenter}}\tau_{\nu_{\hex}}^x
    - J_{\hex} \sin{\varphi} \sum_{\centerhexagon} \mkern9mu \prod_{\mathclap{\nu_{\tri}\in\, \centerhexagon}}\tau_{\nu_{\tri}}^x,
\end{aligned}
\end{equation}
consisting of four- and six-spin \(x\)-interactions as well as magnetic field terms in \(z\)-direction. The four-spin operator involving $\tau_{\nu_{\tri}}^x$ and $\tau_{\nu_{\hex}}^x$ acts on one pseudospin on the hexagonal sublattice and three pseudospins on the triangular sublattice, as illustrated in Fig.~\ref{fig:Dual}. In contrast, the six-spin operator formed by \(\tau_{\nu_{\tri}}^x\)affects six spins on the hexagonal sublattice.
The dual model displays the same energy spectrum as the original model in the correct symmetry sector, except for degeneracies. In the case \(J_{\tri}=0\), the model reduces to the 
color code with only one stabilizer operator up to the non-interacting term \( - J_{\hex} \cos{\varphi}\sum_{\mathrm{Tri}}\tau^z_{\nu_{\hex}}\).
\subsection{Numerical exact diagonalization} \label{sec:ED}
The exponentially growing Hilbert space dimension \(2^N\) for a system with \(N\) spins is reduced by exploiting the model's symmetries. Constraints involve two independent Wilson loop operators and \(\frac{N}{2}-1\) independent \(\hat{A}_s\) symmetries. Thus, in the symmetry sector where \(a_s=1\,\forall s\) and the eigenvalues of each independent non-contractible loop operator are fixed, the effective dimension is reduced to $2^{\frac{N}{2}-1}$.
We consider a system size of \(N=24\) spins (four unit cells) with periodic boundary conditions. For the construction of a suitable basis of the Hilbert space respecting all symmetry constraints, the \texttt{python} package \enquote{QuSpin} is used \cite{QuSpin1}. The \enquote{User\_Basis} class is employed, which provides additional functionality for implementing specific symmetry constraints and customizing corresponding function methods \cite{QuSpin2,quspin_online}. Subsequently, the matrix elements of the Hamiltonian are calculated in this basis. Given the full Hamiltonian matrix, the eigenvalues and eigenvectors are computed numerically with \texttt{linalg.eigsh} from the scipy package, which uses the Implicitly Restarted Lanczos Method \cite{Lanczos,scipy_eigsh}. 

The energy spectrum is identical in the possible topological sectors for both \(\hat{\mathcal{W}}^z\) and \(\hat{\mathcal{W}}^x\) Wilson loop operators. 
\subsection[Phase transition between both Z\textsubscript{2} phases]{Phase transition between both \(\Z_2\) phases} \label{sec:Phase_transitions_between_Z2_phases}
To determine the phase transitions between both \(\Z_2\) topologically ordered phases, order-14 series expansions of the ground-state energy per pseudospin are performed in the dual model \eqref{eq:Hamiltonian_dual_model} using matrix perturbation theory \cite{Oitmaa2006} based on the limits \(\varphi=0\) and \(\varphi = \frac{\pi}{2}\). Both expansions are equivalent, given the symmetric behavior under \mbox{\(\varphi \rightarrow \frac{\pi}{2}-\varphi\)}. 

\begin{figure}[h]
    \centering
    \includegraphics[width=0.9\linewidth]{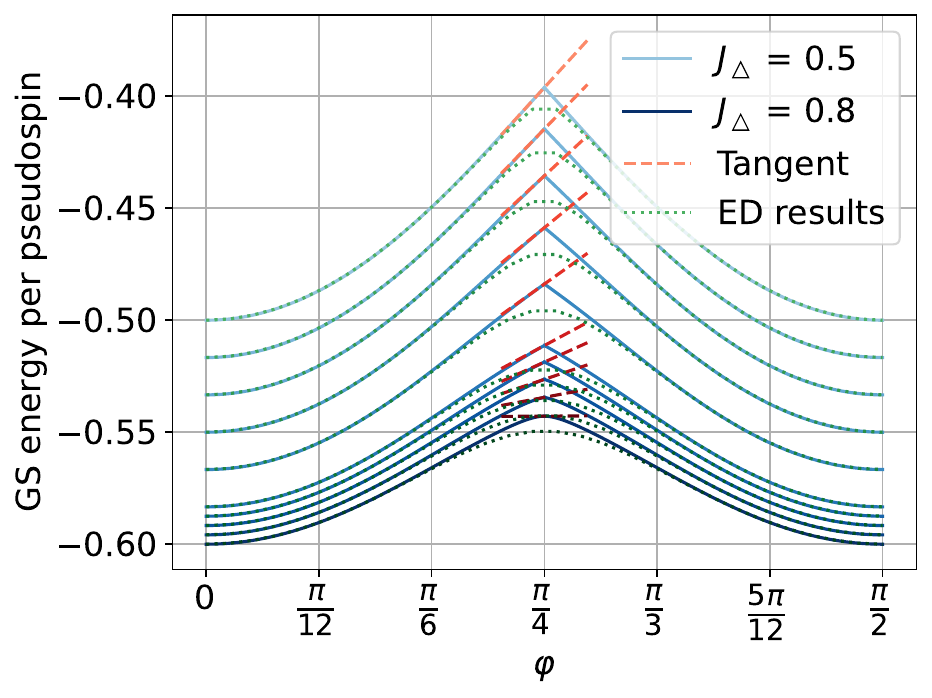}
    \caption[Ground-state energy]{Ground-state energy per pseudospin in the range \mbox{\(J_{\tri}\in[0.5,0.8]\)} determined from matrix perturbation theory. Order-14 perturbation series are displayed in shades of blue, whereas green dotted lines depict numerical exact diagonalization results.
    Results for increasing \(J_{\tri}\) values are shown with increasing opacity. At \mbox{\(\varphi=\frac{\pi}{4}\)}, the tangent of each graph is displayed in red/orange. As \(J_{\tri}\) increases, the slope of the tangent falls off to zero at \(J_{\tri}\approx 0.8\) in order 14.}
    \label{fig:Pert_O14}
\end{figure}

The Hamiltonian \eqref{eq:Hamiltonian_dual_model} is formulated as 
\begin{align} \label{eq:GS_dual_perturbation}
    \frac{\hat{\mathcal{H}}_{\text{dual}}}{\cos{\varphi}\,J_{\hex}} 
    = \hat{\mathcal{H}}_0 + h_0 \hat{V}
\end{align}
with the unperturbed Hamiltonian 
\begin{align}
    \hat{\mathcal{H}}_0 = -\frac{J_{\tri}}{J_{\hex}}\sum_{\text{Hex}} \tau^z_{\nu_{\tri}}-  \sum_{\text{Tri}} \tau^z_{\nu_{\hex}},
\end{align}
perturbation parameter \(h_0=\tan\varphi\), and perturbation 
\begin{align}
    \hat{V} = -\frac{J_{\tri}}{J_{\hex}}\sum_{\centertriangle} \tau_{\nu_{\tri}}^x \prod_{\mathclap{\nu_{\hex} \in \trianglewithoutcenter}}\tau_{\nu_{\hex}}^x - \sum_{\centerhexagon} \mkern9mu \prod_{\mathclap{\nu_{\tri}\in\, \centerhexagon}}\tau_{\nu_{\tri}}^x.
\end{align}
The perturbative series are computed up to order 14 by exploiting hypergraph decompositions \cite{Muehlhauser_2022}.
For each ratio \(J_{\tri}/J_{\hex}\), a different series is obtained due to different energies attributed to local spin flips on the hexagonal sublattice (Hex). 
%The full series for a range of \(J_{\tri}\) values are provided in an \hl{additional data file}. 
Fig.~\ref{fig:Pert_O14} shows the ground-state energy per pseudospin as perturbative order-14 expansions from \(\varphi = 0\) and \(\varphi = \frac{\pi}{2}\) for \(J_{\tri}\in [0.5,0.8]\). Series for a given value of \(J_{\tri}\) intersect at precisely \(\varphi = \frac{\pi}{4}\). The intersection angle is finite for most \(J_{\tri}\in [0.5,0.8)\) values, decreasing to zero for \(J_{\tri} \approx 0.8\) in order 14.
The perturbative results are consistent with exact diagonalization.

\subsection[Phase transition between ZZZ and Z\textsubscript{2} phases]{Phase transition between \(\ZZZ\) and \(\Z_2\) phases}
The phase transition between the \(\ZZZ\) and \(\Z_2\) topologically ordered phases is qualitatively determined from numerical exact diagonalization and by series expansions of the ground-state energies. Both results are presented and compared in the following.
\subsubsection{Phase transition based on exact diagonalization}
\begin{figure}[h]
    \centering
    \begin{minipage}{0.48\linewidth}
        \centering
        \includegraphics[width=\linewidth]{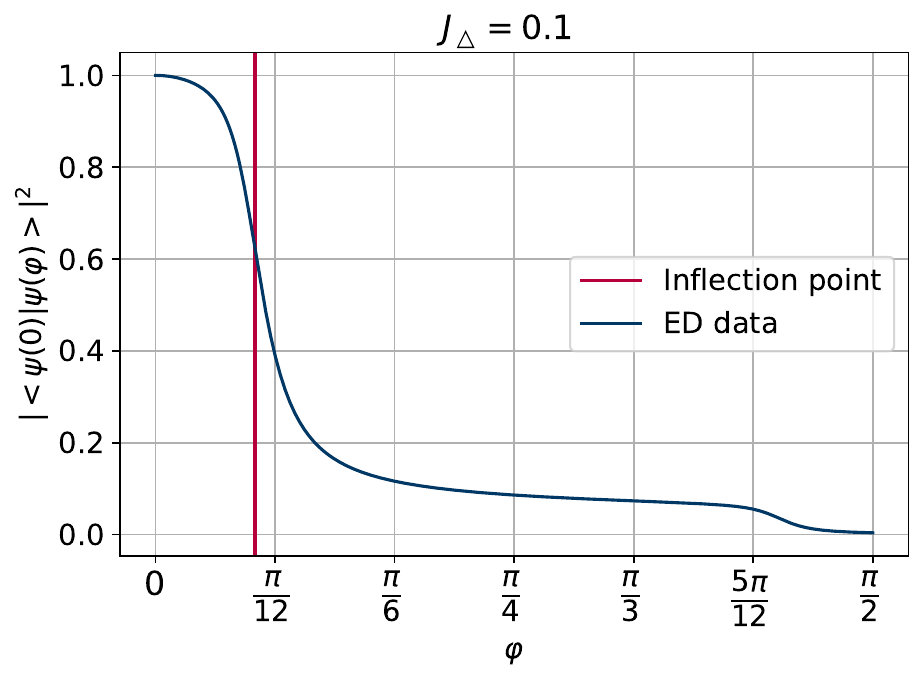}
    \end{minipage}
    \hfill
    \begin{minipage}{0.48\linewidth}
        \centering
        \includegraphics[width=\linewidth]{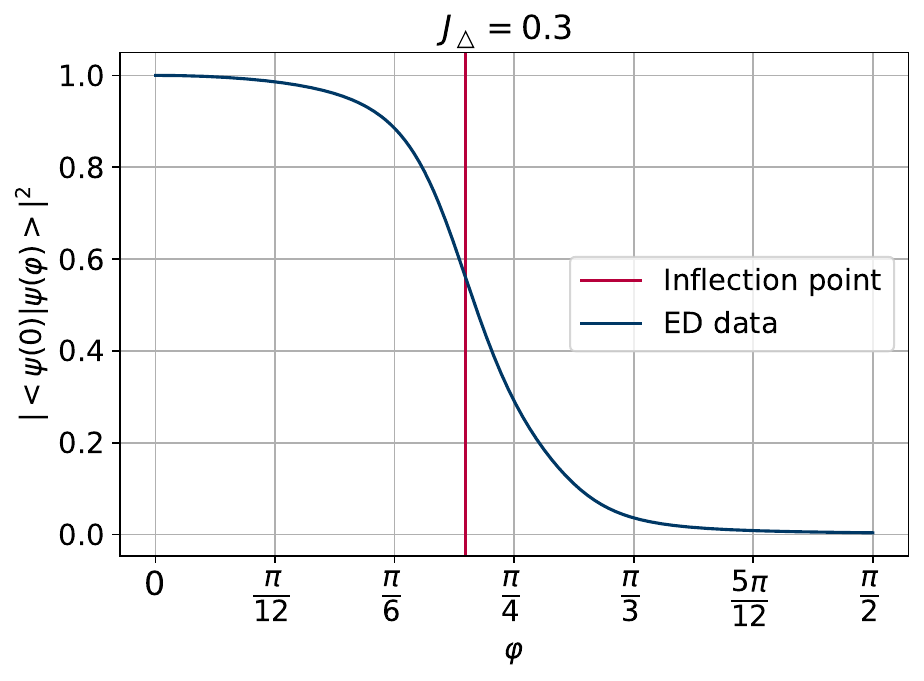}
    \end{minipage}
    \caption[Overlap]{Overlap \(\left|\braket{\psi(0) | \psi(\varphi)}\right|^2\) of the ground states calculated from exact diagonalization (blue) and numerically determined inflection point (red). \textit{Left}: Curve for \(J_{\tri}=0.1\). \textit{Right}: Curve for \(J_{\tri}=0.3\).}
    \label{fig:Inflection_points}
\end{figure}

\begin{figure}
    \centering
    \includegraphics[width=0.8\linewidth]{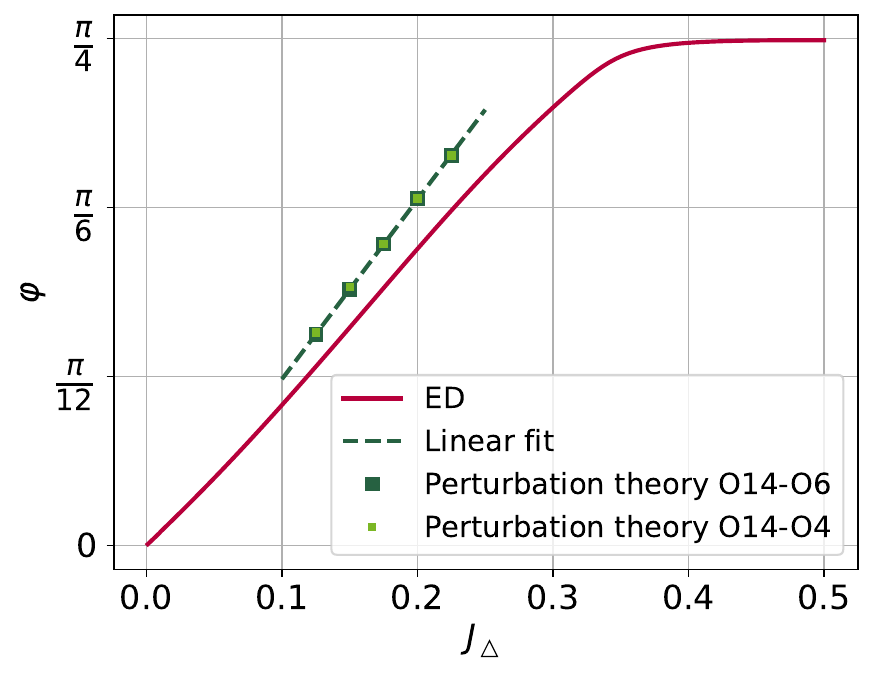}
    \caption[Phase transition line between ZZZ and Z2 phases as determined from exact diagonalization and perturbation theory.]{Phase transition line between \(\ZZZ\) and \(\Z_2\) phases as determined from exact diagonalization (red line) and perturbation theory (green squares). Light green squares refer to order-4 expansions about \(J_{\tri}=0\) and order-14 expansions about \(\varphi = 0\), while dark green squares correspond to order-6 expansions about \(J_{\tri}=0\) and order-14 expansions about \(\varphi = 0\). The deviation is negligible. Further, reducing the order-14 expansion to include only orders up to order 10 does not noticeably influence the intersection points. The green dashed line represents a linear fit to the perturbative O14-O6 points with \mbox{\(\varphi(J_{\tri}) = (2.784 \pm 0.018)J_{\tri} + (-0.021\pm 0.004)\)}.}
    \label{fig:Phase_transitions}
\end{figure}

Exact diagonalization in a system comprising 24 spins is performed according to sec.~\ref{sec:ED}. To assess the point of phase transition, we first compute the eigenvectors \(\ket{\psi(\varphi)}\) corresponding to the ground-state energy for fixed \(J_{\tri}\). We then calculate the overlap of \(\ket{\psi(\varphi)}\) for a given value of \(\varphi\) with the ground state at \(\varphi = 0\),
\begin{align}
    \left|\braket{\psi(0) | \psi(\varphi)}\right|^2,
\end{align}
as a function of \(\varphi\). By numerically computing the second derivative, we determine the first point of inflection of each curve. As an example, the curves for \(J_{\tri}=0.1\) and \(J_{\tri}=0.3\) are shown in Fig.~\ref{fig:Inflection_points} to illustrate the procedure.
To accumulate a sufficient number of data points, we consider \(100\) \(J_{\tri}\) values in the range \(J_{\tri}\in [0.01, 0.5]\), each involving 300 \(\varphi\) values in the range \(\varphi \in [0,\frac{\pi}{2}]\). For \mbox{\(J_{\tri}<0.01\)}, we choose a smaller range  \(\varphi \in [0,\frac{\pi}{12}]\) and smaller step size.
Plotting the so-determined inflection points, we obtain the phase transition curve as shown in red in Fig.~\ref{fig:Phase_transitions}. 
The results obtained from exact diagonalization obviously suffer from finite-size effects due to the limited system size and large unit cell. Therefore, the exact phase transition lines in the thermodynamic limit are not expected to agree quantitatively.
However, finite-size effects are expected to be small due to the first-order nature of the phase transitions.
\subsubsection{Phase transition based on series expansion}
An alternative approach not restricted by finite-size effects involves calculating the intersection points of perturbative ground-state energy series expansions about the limits \(\varphi = 0 \) and \(J_{\tri}=0\).

The order-14 series of the ground-state energy per pseudospin as an expansion in the perturbation parameter \(h_0=\tan\varphi\) about \(\varphi=0\) is determined as described in sec.~\ref{sec:Phase_transitions_between_Z2_phases}. The smaller \(J_{\tri}\), the smaller the radius of convergence of the series.

Further, we make use of the ground-state energy series expansion about the limit \(J_{\tri}=0\). Up to order 6, the series \eqref{eq::ham_eff_supp} with \(s_{\dodec}^{(x)}=1\) is equivalent to that of two decoupled quantum Baxter-Wu models \cite{Bax_Wu_original}, since the non-trivial particle statistics are still irrelevant.
This can be seen by considering the model 
\begin{equation}
\hat{\mathcal{H}}_0-J_{\tri}\sum_{\tri} \hat{B}_{\tri}^{(\alpha)} 
\end{equation}
with \mbox{\(\hat{\mathcal{H}}_0=-J_s\sum_s \hat{A}_s - J_{\hex}\sum_{\hex}\hat{B}_{\hex}\)}
in the symmetry sector \(s_{\dodec}=+1\). By introducing pseudospins \(\tau^{\alpha}_{\nu_{\hex}}\) in the center of the hexagonal plaquettes and identifying the eigenvalues of  \(\tau^{z}_{\nu_{\hex}}\) with \(b^{(\alpha)}_{\hex}=\pm 1\) while the other \(\hat{B}_{\hex}^{(\alpha ')}\) remains a symmetry and \(b^{(\alpha ')}_{\hex}=+1\) is chosen, one obtains a dual model. This model is then given by one copy of the quantum Baxter-Wu model 
\begin{equation}
\sum_{\text{Tri}} \tau^{z}_{\nu_{\hex}} - \sum_{\trianglewithoutcenter} \prod_{\nu_{\hex} \in \trianglewithoutcenter} \tau^{x}_{\nu_{\hex}}
\end{equation}
with three-spin interactions on the triangular lattice (Tri). The perturbative series of the ground-state energy for \( \hat{\mathcal{H}}_0 - J_{\hex}\sum_{\hex}\hat{B}_{\hex}\) is therefore identical to two quantum Baxter-Wu models up to order 6, since there are no processes in which the nontrivial statistic between \(B^{x}\) and \(B^{y}\) particles is relevant up to this order.
Series expansions to higher orders taking into account the non-trivial statistic might be possible with more involved methods \cite{Muehlhauser_2024}.

We then determine the intersection points of the \mbox{order-14} series about \(\varphi = 0 \) and the order-6 series about \mbox{\(J_{\tri}=0\)}. Due to convergence issues, only in the range \mbox{\(J_{\tri}\in[0.125,0.25]\)} do we find reliable intersections. The so-determined intersection points are displayed in green in Fig.~\ref{fig:Phase_transitions}.

The results obtained from exact diagonalization and perturbative series expansions agree qualitatively in that the shape of the phase transition curve is approximately linear in the parameter regime \(J_{\tri}\in[0.125,0.25]\). However, the perturbatively determined intersection points are shifted upwards compared to the ED results, and differ in the slope of the linear incline. These differences are expected due to the finite system size of the numerical exact diagonalization calculation and convergence problems of the series expansions. 
%\bibliography{bibliography_supplemental}

\end{document}